\documentclass[showpacs,preprint,amsmath,amssymb,aps,prd,eqsecnum]{revtex4}
\usepackage{graphicx}
\usepackage{indentfirst} 
\begin{document}

\title{Black Objects and Hoop Conjecture in Five-dimensional Space-time}

\author{Yuta Yamada
\footnote{ m1m08a26@info.oit.ac.jp} and 
Hisa-aki Shinkai
\footnote{shinkai@is.oit.ac.jp}}
\address{Faculty of Information Science and Technology, Osaka Institute of Technology,\\ 1-79-1 Kitayama, Hirakata, Osaka 573-0196, Japan~ }

\date{\today}

\begin{abstract}
We numerically investigated the sequences of initial data of thin spindle and thin ring in five-dimensional space-time in the context of the cosmic censorship conjecture. We modeled the matter in non-rotating homogeneous spheroidal or toroidal configurations under the momentarily static assumption, solved the Hamiltonian constraint equation, and searched the apparent horizons. We discussed when $S^3$ (black hole) or $S^1 \times S^2$ (black ring) horizons (``black objects") are formed. By monitoring the location of the maximum Kretchmann invariant, an appearance of `naked singularity' or `naked ring' under special situation is suggested. We also discuss the validity of the {\it hyper-hoop} conjecture using minimum {\it area} around the object, and show that the appearance of the ring horizon does not match with this hoop.
\end{abstract}

\maketitle
\clearpage

\section{Introduction}

In general relativity, there are two famous conjectures concerning the gravitational collapse.
One is the cosmic censorship conjecture \cite{Penrose} which states collapse driven singularities will always be clothed by an event horizon and hence can never be visible from the outside.
The other is the hoop conjecture \cite{thorne} which states that black-holes will form when and only when a mass $M$ gets compacted into a region whose circumference $C$ in every direction is $C\leq4\pi M$.
These two conjectures have been extensively studied in various methods, among them we believe the numerical
works by Shapiro and Teukolsky \cite{shapiro} showed the most exciting results; (a tendency of) the appearance of a naked singularity.  This was reported from the fully relativistic time evolution of
collisionless particles in a highly prolate initial shape; and the results of time evolutions are agree
with the predictions of the sequence of their initial data \cite{nakamura}.

In recent years, on the other hand, gravitation in higher-dimensional spacetime is getting a lot of attention.  This is from an attempt to unify fundamental forces including gravity at TeV scale, and if so,
it is suggested that small black-holes might be produced at the CERN Large Hadron Collider (LHC).
The LHC experiments are expected to validate several higher-dimensional gravitational models.  In such an exciting situation, the thoretical interests are also in the general discussion of black-hole structures.
Our discussion is one of them: in what circumstances black-holes are formed?

New features of higher-dimensional black-holes and black-objects are reported due to additional physical freedoms.
The four-dimensional black-holes are known to be $S^2$ from the topological theorem.
Also in the asymptotically flat and stationary space-time, four-dimensional black-holes are known to be the Kerr black-hole from the uniqueness theorem.
On the other hand, in higher-dimensional spacetime, quite rich structures are available, such as a torus black-hole (``black ring") with $S^1 \times S^2$ horizon \cite{Emparan, blackring2} or black Saturn \cite{Elvang}, black di-ring\cite{Krishnan,izumi} (see a review \cite{Emparan2} for references).
The uniqueness theorem of axisymmetric spacetime in higher-dimension is known to be violated.

So far, the black-hole studies in higher-dimensional spacetime are mainly proceeding using analytic stationary solutions.
There are also many numerical attempts to seek the higher-dimensional 
black-hole structures; e.g. collider-oriented dynamical features \cite{EardleyGiddings, yoshino}, 
a new stationary solution sequence \cite{kudoh}, 
(here we selected the works with asymptotically flat spacetime).  
However, fully relativistic dynamical features, such as the formation processes, stabilities, and late-time fate of the black-objects are left unknown.
We plan to investigate such dynamical processes numerically, and this is the first report on the constructions of the sequences of initial data for time evolution.

The hoop conjecture tries to denote ``if" and ``only if"
conditions for the formation of the horizon in the process of
gravitational collapse.
The ``only if" part of the statement would be replaced with
the so-called Gibbons-Penrose isoperimetric inequality
\cite{Gibbons-Penrose}, $M \geq \sqrt{A/16 \pi}$, where $M$ is the
total mass and $A$ is the area of the trapped surface.
This inequality is based on the cosmic censorship conjecture, so that
its proof or disproof is the important issue (see a precise formulation in \cite{Gibbons} and a recent review 
\cite{Mars09}).

The higher dimensional versions of the hoop conjecture and the 
isoperimetric inequality
have been discussed so far \cite{ida, BFL2004, 
GibbonsHolzegel2006, Seno2008}.
While there are differences in their coefficiencies, the hoop conjecture
in $D$-dimensional spacetime would be basically expressed as follows: 
a black-hole
with horizons form when and only when a mass $M$ gets compacted into a 
region
whose $(D-3)$-dimensional area $V_{D-3}$ in every direction is
\begin{equation}
V_{D-3} \leq G_D M, \label{1.1}
\end{equation}
where $G_D$ is the gravitational constant in $D$-dimensional theory of 
gravity.
Here $V_{D-3}$ means the volume of $(D-3)$-dimensional
closed submanifold of a spacelike hypersurface. That is, the hoop $C$ in
four-dimensional space-time is replaced with the {\it hyper-hoop} 
$V_{D-3}$;
if $D=5$, then the {\it hyper-hoop} would be an area $V_2$.
However, in five-dimensional spacetime, black-holes are not restricted to 
have a simply-connected horizon,
therefore the applicabilities of the hyper-hoop and the 
isoperimetric inequality
to various black-objects are left unknown.
The validity of (\ref{1.1}) was investigated in several idealized 
models by Ida and Nakao \cite{ida} and
Yoo et al. \cite{yoo}, who solved momentarily static, conformally flat,
five-dimensional axisymmetric homogeneous spheroidal matter and
$\delta$-function type ring matter. Our purpose is to investigate the 
generality of the hyper-hoop conjecture
and the cosmic censorship conjecture in more general situations. 

In this article, we present two kinds of initial data; spheroidal and 
toroidal matter configurations.
We solve the Hamiltonian constraint equation numerically, and then 
search apparent horizons.
This study is the generalization of \cite{ida} and \cite{yoo}; we 
reproduce their results as
our code checks, and present also finite-sized ring cases.
The definition of  the hyper-hoop is not yet definitely given in the community, so that we propose 
to define the hyper-hoop
as a local minimum of the area by solving the Euler-Lagrange type 
equation.

This article is organized as follows.
In the next section, we explain how to set initial data for five-dimensional space-time and how to search $S^3$ and $S^1 \times S^2$ apparent horizons and hoops.
In Sec. III, we show numerical results.
The final section is devoted to the summary and discussion.
We use the unit $c=1$ and $G_5=1$, where $c$ is the speed of light, $G_5$ is the gravitational constant of the five-dimensional spacetime.

\section{Basic Equations \& Numerical Issues}
\subsection{The Hamiltonian constraint equation}

We consider the initial data sequences on a four-dimensional space like hypersurface.
A solution of the Einstein equations is obtained by solving the Hamiltonian constraint equation
if we assume the moment of time symmetry.
We apply the standard conformal approach\cite{york} to obtain the four-metric $\gamma_{ij}$.
As was discussed in \cite{shinkai}, in $4+1$ space-time decomposition, the equations would be simplified with a conformal transformation,
\begin{equation}
\gamma_{ij} = \psi^2\hat{\gamma}_{ij}, 
\end{equation}
where $\hat{\gamma}_{ij}$ is the trial base metric which we assume conformally flat,
\begin{equation}
ds^2=\hat{\gamma}_{ij}dx^idx^j=dx^2+dy^2+dz^2+dw^2.
\end{equation}

The Hamiltonian constraint equation, then, becomes
\begin{equation}
\hat{\Delta}\psi = -4\pi^2G_{5} \rho \label{Hamiltonian},
\end{equation}
where $\rho$ is the effective Newtonian mass density, $G_{5}$ is the gravitational constant in five-dimensional theory of gravity.
We numerically solve Eq.(\ref{Hamiltonian}) 
in the upper-half coordinate region ($x\ge 0,\;y\ge 0,\;z\ge 0,\;w\ge 0$) with 
setting the boundary conditions as 
\begin{equation}
\nabla\psi = 0 \;\;\;\mbox{(at inner boundaries)},
\end{equation}
and,
\begin{equation}
\psi = 1+\frac{M_{ADM}}{r^2}\;\;\;\mbox{(at outer boundaries)},\label{boundary}
\end{equation}
where
\begin{equation}
r=\sqrt{x^2+y^2+z^2+w^2} \label{xyzw}
\end{equation}
and $M_{ADM}$ can be interpreted as the ADM mass of the matter.
Practically, the boundary condition, (\ref{boundary}), is replaced with
\begin{equation}
(\psi-1)r^2=\mbox{const}.
\end{equation}
and we apply
\begin{equation}
\frac{\partial}{\partial x^i}\left[(\psi-1)r^2\right]=0
\end{equation}
on the outer edge of our numerical grid.
The ADM mass $M_{ADM}$, then, is evaluated from Eq.(\ref{boundary}).
\begin{figure}[!htbp]
\centering
\includegraphics[width=8cm, clip]{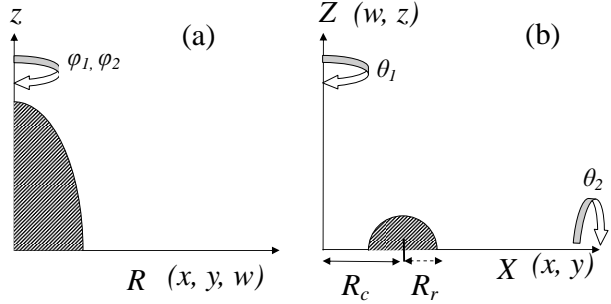}
\caption{Axis of symmetry of our models: (a) spheroidal (spindle) configuration, and (b) toroidal configuration. We consider the matter with uniform density. We adopt the coordinate as Eq.(\ref{metric}) for the case (a), while we use Eq.(\ref{metric2}) for the case (b).}
\end{figure}

As is described below, we consider two models of the matter distribution : spheroidal and toroidal configurations.
By assuming the axis of symmetry, both are reduced to effectively two-dimensional problems (Fig.1).
For the spheroidal matter $\left[\right.$Fig.1(a)$\left.\right]$, we use the metric
\begin{equation}
ds^2=\psi(R,z)^2\left[dR^2+R^2(d\varphi^2_1+\sin^2\varphi_1d\varphi^2_2)+dz^2\right] \label{metric}
\end{equation}
where
\begin{eqnarray}
R&=&\sqrt{x^2+y^2+z^2},\;\; 
\varphi_1=\tan^{-1}\left(\frac{w}{\sqrt{x^2+y^2}}\right),\;\;
\mbox{and}\;\;
\varphi_2=\tan^{-1}\left(\frac{y}{x}\right).\nonumber
\end{eqnarray}
For the toroidal case $\left[\right.$Fig.1(b)$\left.\right]$, on the other hand, we use the metric
\begin{equation}
ds^2=\psi(X,Z)^2(dX^2+dZ^2+X^2d\vartheta_1+Z^2d\vartheta_2) \label{metric2}
\end{equation}
where
\begin{eqnarray}
X &=& \sqrt{x^2+y^2}, \;\; 
Z = \sqrt{z^2+w^2}, \nonumber \\
\vartheta_1 &=& \tan^{-1}\left(\frac{y}{x}\right),\;\;\mbox{and}\;\; 
\vartheta_2 = \tan^{-1}\left(\frac{z}{w}\right).\nonumber
\end{eqnarray}
By assuming $\varphi_1$ and $\varphi_2$ ($\vartheta_1$ and $\vartheta_2$ for the toroidal case) are the angle around the axis of symmetry, then the Hamiltonian constraint equation, (\ref{Hamiltonian}), effectively becomes
\begin{equation}
\frac{\partial^2\psi}{\partial R^2}+\frac{2}{R}\frac{\partial\psi}{\partial R}+\frac{\partial^2\psi}{\partial z^2}=-4\pi^2G_{5} \rho, \label{HCE}
\end{equation}
and
\begin{equation}
\frac{1}{X}\frac{\partial}{\partial X}\left(X\frac{\partial\psi}{\partial X}\right)+\frac{1}{Z}\frac{\partial}{\partial Z}\left(Z\frac{\partial\psi}{\partial Z}\right)= -4\pi^2G_{5} \rho, \label{HCE2}
\end{equation}
respectively.
We solve (\ref{HCE}) and (\ref{HCE2}) using normal the successive over-relaxation(SOR) method with red-black ordering.
We use $500^2$ grids for the range $(R, z)$ or $(X, Z) =\left[0,10\right]$ with the tolerance $10^{-6}$ for $\psi$ for solving Eqs.(\ref{HCE}) and (\ref{HCE2}).
The presenting results are the sequences of the constant $M_{ADM}$ within the error O($10^{-2}$).

\subsection{Matter distributions}

We model the matter by non-rotating homogeneous spheroidal and toroidal configurations with effective Newtonian uniform mass density.
Our first model is the cases with homogeneous spheroidal matter, which are expressed as
\begin{equation}
\frac{x^2}{a^2}+\frac{y^2}{a^2}+\frac{w^2}{a^2}+\frac{z^2}{b^2}\leq 1,\label{spindle}
\end{equation}
where $a$ and $b$ are parameters.
This is the $4+1$ dimensional version of the earlier study of Nakamura et al. \cite{nakamura}, and also the numerical reproduction of Ida-Nakao \cite{ida} and Yoo et al. \cite{yoo}.

The second is the cases with homogeneous toroidal matter configurations, described as
\begin{equation}
\left(\sqrt{x^2+y^2}-R_c\right)^2+\left(\sqrt{w^2+z^2}\right)^2\leq R_r^2, \label{toroidal}
\end{equation}
where $R_c$ is the circle radius of torus, and $R_r$ is the ring radius $\left[\right.$Fig.1(b)$\left.\right]$.
This case is motivated from the ``black ring" solution \cite{Emparan} though not including any rotations of matter nor of the spacetime.
Nevertheless we consider this is the first step for toroidal configuration, since this is the generalization of \cite{ida} to the finite-sized matter cases.

\subsection{Kretchmann invariant}

After obtaining the initial data, we evaluate the Kretchmann invariant,
\begin{equation}
{\cal I}^{(4)} = R_{abcd}R^{abcd}\label{invariant},
\end{equation}
where $R_{abcd}$ is the four-dimensional Riemann tensor, in order to measure the strength of gravity.
This is most easily evaluated in Cartesian coordinates as
\begin{eqnarray}
{\cal I}^{(4)}
        &=& 
 16\sum_{i\neq j} 
 \left[ 2\,\left( {\frac {\partial \psi}{\partial x^i}}  \right) 
           \left( {\frac {\partial \psi}{\partial x^j}}  \right)
      -\psi {\frac {\partial^{2}\psi}{\partial x^i \partial x^j}} \right] ^{2} 
 \nonumber \\&& 
 + 8 \sum_{i\neq j} \left[\left( {\frac {\partial \psi }{\partial x^i}} \right) ^2-\left( {\frac {\partial \psi }{\partial x^j}}\right) ^2\right]^2 
 + 4 \psi^2 \sum_{i\neq j} \left[ \frac {\partial ^{2}\psi}{\partial {x^i}^{2}} +\frac {\partial ^{2}\psi}{\partial {x^j}^{2}} \right]^2
 \nonumber \\&& 
 + 8 \psi
 \left[ \sum_{i} \left( {\frac {\partial \psi }{\partial x^i}} \right) ^2 \right]
 \left[ \sum_{i}  {\frac {\partial^2 \psi }{\partial {x^i}^2}}  \right]
 -32 \psi \sum_{i} \left( {\frac {\partial \psi }{\partial x^i}} \right) ^2  
 \left( {\frac {\partial^2 \psi}{\partial {x^i}^2}} \right).\nonumber
\end{eqnarray}

\subsection{Apparent Horizons}

For investigating the validity of the censorship conjecture and hyper-hoop conjecture, we search the existence of apparent horizons.
An apparent horizon is defined as a marginally outer trapped surface, 
and the existence of the apparent horizon is the sufficient condition for the existence of the event horizon.
On the four-dimensional spacelike hypersurface, an apparent horizon is a three-dimensional closed marginal surface.

In order to locate the apparent horizon for the spheroidal configurations, after obtained the solution of (\ref{HCE}), we transform the coordinate 
from $(R,z)$ to $(r, \theta)$, using
\begin{eqnarray}
r&=&\sqrt{R^2+z^2},\\
\theta&=&\tan^{-1}\left(\frac{R}{z}\right),
\end{eqnarray}
and search the apparent horizon on the $R$-$z$ section \cite{ida, yoo}.
The location of the apparent horizon, $r_M(\theta)$, is identified by solving
\begin{eqnarray}
\ddot{r}_M-\frac{4\dot{r}^2_M}{r_M}-3r_M+\frac{r^2_M+\dot{r}^2_M}{r_M}\left[\frac{2\dot{r}_M}{r_M}\cot\theta-\frac{3}{\psi}(\dot{r}_M\sin\theta\right. \nonumber \\
\left.+r_M\cos\theta)\frac{\partial\psi}{\partial z}+\frac{3}{\psi}(\dot{r}_M\cos\theta-r_M\sin\theta)\frac{\partial\psi}{\partial R}\right] = 0, 
\label{AH}
\end{eqnarray}
where dot denotes $\theta$-derivative.
We solve (\ref{AH}) for $r_M(\theta)$ using Runge-Kutta method starting on the $z$-axis ($\theta=0$) with a trial value $r=r_0$ and integrate to $\theta=\pi/2$, with interpolating the coefficients $\psi$ and $\displaystyle\frac{\partial \psi}{\partial x^i}$ from the data on the grid points.
We apply the symmetric boundary condition on the both ends.
If there is no solution satisfying both boundary conditions, we judge there is no horizon.

For toroidal cases, we transform the coordinate from $(X,Z)$ to $(r, \phi)$, using
\begin{eqnarray}
r&=&\sqrt{X^2+Z^2}, \;\; \mbox{and}\;\;
\phi=\tan^{-1}\left(\frac{Z}{X}\right).
\end{eqnarray}
The location of the apparent horizon, $r_m(\phi)$, is then identified by solving
\begin{eqnarray}
\ddot{r_m}-4\frac{\dot{r_m}^2}{r_m}-3r_m-\frac{r^2_m+\dot{r_m}^2}{r_m}\left[2\frac{\dot{r_m}}{r_m}\cot(2\phi)-\frac{3}{\psi}(\dot{r_m}\sin\phi \right. \nonumber \\
\left.+r\cos\phi)\frac{\partial\psi}{\partial X}+\frac{3}{\psi}(\dot{r_m}\cos\phi-r_m\sin\phi)\frac{\partial\psi}{\partial Z}\right] = 0, \label{AH2}
\end{eqnarray}
with the symmetric boundary condition $\dot{r}=0$ at both  $\phi=0$ and $\pi/2$.
When the matter is in torus shape, an additional $S^1 \times S^2$ apparent(ring horizon) horizon may 
exist. 
In order to find a ring horizon, we adopt the coordinate as
\begin{eqnarray}
r = \sqrt{(X-R_c)^2 + Z^2},\;\;\mbox{and}\;\;
\xi=\tan^{-1}\left(\frac{Z}{X-R_c}\right). \label{defxi}
\end{eqnarray}
This marginal surface is obtained by solving the equation for 
$r(\xi)$, 
\begin{eqnarray}
\ddot{r_m}-\frac{3\dot{r_m}^2}{r_m}-2r_m-\frac{r^2_m+\dot{r_m}^2}{r_m}\times\left[\frac{\dot{r_m}\sin\xi+r_m\cos\xi}{r_m\cos\xi+R_c} \right. \nonumber \\
-\frac{\dot{r_m}}{r_m}\cot\xi+\frac{3}{\psi}(\dot{r_m}\sin\xi+r\cos\xi)\frac{\partial\psi}{\partial x} 
\left.-\frac{3}{\psi}(\dot{r_m}\cos\xi-r\sin\xi)\frac{\partial\psi}{\partial z}\right] = 0,  \label{AH3}
\end{eqnarray}
where dot denots $\xi$-derivative, with the symmetric boundary condition on the both ends at $\xi=0$ and $\pi$.

\subsection{Area of horizons}

From the obtained sequence of initial data, we calculate the surface area $A_3$ of the apparent horizons.
If the obtained horizon is spheroidal configuration, the surface area of the horizon, $A_3$, becomes
\begin{equation}
A_3^{(S)} = 8\pi\int^{\pi/2}_{0}\psi^3r_{M}^2\sin^2\theta\sqrt{\dot{r_M}^2+r_{M}^2}\,d\theta,
\end{equation}
where dot denotes $\theta$-derivative.
As for the toroidal cases, the surface area of $S^3$ and $S^1 \times S^2$ apparent horizons become
\begin{eqnarray}
A_3^{(T1)}&=&4\pi^2\int^{\pi/2}_{0}\psi^3r_{m}^2\cos\phi\sin\phi\sqrt{\dot{r_m}^2+r_{m}^2}\,d\phi,\label{V3T1}
\end{eqnarray}
and
\begin{eqnarray}
A_3^{(T2)}&=&4\pi^2\int^{\pi}_{0}\psi^3(R_c+r_{m}\cos\xi)r_{m}\sin\xi\sqrt{\dot{r_m}^2+r_{m}^2}\,d\xi, \label{V3T2}
\end{eqnarray}
where dot denotes $\phi$-derivative and $\xi$-derivative, respectively.

\subsection{Hyper-Hoop}
We also calculate hyper-hoop for five-dimensional hoop-conjecture which is defined by two-dimensional area. We try to verify the necessary condition of the black-hole formation examined in \cite{yoo},
\begin{equation}
V_2 \leq \frac{\pi}{2}16\pi G_5M. \label{sufficient_condition}
\end{equation}
However, the definition of $V_2$ is not so far defined apparently.
We, therefore, propose to define the hoop $V_2$ as a surrounding two-dimensional area which satisfies the local minimum area condition,
\begin{equation}
\delta V_2 = 0.\label{henbun}
\end{equation}
When the area of the space-time outside the matter is expressed by a coordinate $r$, then Eq.(\ref{henbun}) leads to the Euler-Lagrange type equation for $V_2(r,\dot{r})$.

For the spheroidal configuration, we express the area $V_2$ using $r = r_h(\theta)$ as
\begin{equation}
V_2^{(A)}=4\pi\int^{\pi/2}_{0}\psi^2\sqrt{\dot{r_h}^2+r_h^2}r_h\sin\theta\,
d\theta,
\end{equation}
or
\begin{equation}
V_2^{(B)}=4\pi\int^{\pi/2}_{0}\psi^2\sqrt{\dot{r_h}^2+r_h^2}r_h\cos\theta\, d\theta,
\end{equation}
where dot denotes $\theta$-derivative.
$V_2^{(A)}$ expresses the surface area which is obtained by rotating respect to the $z$-axis, while $V_2^{(B)}$ is the one with $R$-axis rotation. 
Then the hyper-hoop $V_2^{(A)}$ is derived by
\begin{eqnarray}
\ddot{r_h}-\frac{3\dot{r_h}^2}{r_h}-2r_h+\frac{r^2_h+\dot{r_h}^2}{r_h}\left[\frac{\dot{r_h}}{r_h}\cot\theta-\frac{2}{\psi}(\dot{r_h}\sin\theta \right.\nonumber \\
\left.+r_h\cos\theta)\frac{\partial \psi}{\partial z}-\frac{2}{\psi}(r_h\sin\theta-\dot{r_h}\cos\theta)\frac{\partial \psi}{\partial R}\right]=0,
\label{minV}
\end{eqnarray}
while the hyper-hoop $V_2^{(B)}$ is by
\begin{eqnarray}
\ddot{r_h}-\frac{3\dot{r_h}^2}{r_h}-2r_h-\frac{r^2_h+\dot{r_h}^2}{r_h}\left[\frac{\dot{r_h}}{r_h}\tan\theta+\frac{2}{\psi}(r_h\sin\theta \right.\nonumber \\
\left.-\dot{r_h}\cos\theta)\frac{\partial \psi}{\partial R}+\frac{2}{\psi}(r_h\cos\theta+\dot{r_h}\sin\theta)\frac{\partial \psi}{\partial z}\right]=0.
\label{minV2}
\end{eqnarray}
We search the location of the minimum $V_2$ by solving (\ref{minV}) and (\ref{minV2}), applying the same technique and the boundary conditions with those of horizons.

For the toroidal cases, the hoop is expressed using $r = r_h({\it \phi})$ as
\begin{equation}
V_2^{(C)}=4\pi\int^{\pi/2}_{0}\psi^2\sqrt{\dot{r_h}^2+r_h^2}r_h\cos\phi\, d\phi,
\end{equation}
or
\begin{equation}
V_2^{(D)}=4\pi\int^{\pi/2}_{0}\psi^2\sqrt{\dot{r_h}^2+r_h^2}r_h\sin\phi\, d\phi.
\end{equation}
$V_2^{(C)}$ expresses the surface area which is obtained by rotating respect to the $Z$-axis, while $V_2^{(D)}$ is the one with $X$-axis rotation. 
Then, the minimum $V_2^{(C)}$ satisfies the equation
\begin{eqnarray}
\ddot{r_h}-\frac{3\dot{r_h}^2}{r_h}-2r_h+\frac{r^2_h+\dot{r_h}^2}{r_h}\left[\frac{\dot{r_h}}{r_h}\cot\phi-\frac{2}{\psi}(\dot{r_h}\sin\phi \right.\nonumber \\
\left.+r_h\cos\phi)\frac{\partial \psi}{\partial X}-\frac{2}{\psi}(r_h\sin\phi-\dot{r_h}\cos\phi)\frac{\partial \psi}{\partial Z}\right]=0,
\label{minVC}
\end{eqnarray}
and $V_2^{(D)}$ satisfies
\begin{eqnarray}
\ddot{r_h}-\frac{3\dot{r_h}^2}{r_h}-2r_h-\frac{r^2_h+\dot{r_h}^2}{r_h}\left[\frac{\dot{r_h}}{r_h}\tan\phi+\frac{2}{\psi}(r_h\sin\phi \right.\nonumber \\
\left.-\dot{r_h}\cos\phi)\frac{\partial \psi}{\partial X}+\frac{2}{\psi}(r_h\cos\phi+\dot{r_h}\sin\phi)\frac{\partial \psi}{\partial Z}\right]=0.
\label{minVD}
\end{eqnarray}
We also calculate hyper-hoop with $S^1\times S^1$ topology for the toroidal cases, $V_2^{(E)}$,
\begin{equation}
V_2^{(E)}=2\pi\int^{\pi}_{0}\psi^2\sqrt{\dot{r_h}^2+r_h^2}(r_h\cos\xi+R_c)\, d\xi.
\end{equation}
The minimum $V_2^{(E)}$ satisfies the equation,
\begin{eqnarray}
\ddot{r_h}-\frac{3\dot{r_h}^2}{r_h}-2r_h-\frac{r^2_h+\dot{r_h}^2}{r_h}\left[\frac{-R_c+\dot{r_h}\sin\xi}{R_c+r_h\cos\xi}+\frac{2}{\psi}(\dot{r_h}\sin\xi\right.\nonumber \\
\left.+r_h\cos\xi)\frac{\partial \psi}{\partial X}+\frac{2}{\psi}(r_h\sin\xi-\dot{r_h}\cos\xi)\frac{\partial \psi}{\partial Z}\right]=0.
\end{eqnarray}

\section{Numerical results}\label{results}
\subsection{Spheroidal configurations}
First, we show the cases with spheroidal matter configurations.
In Figure 2, we display matter distributions and the
shape of the apparent horizon (if it exists). 
When the matter is spherical, $a=b$ $\left[\right.$the cases of (a), (d) in Fig.2$\left.\right]$, the horizon is also spherically symmetric and locates at the Schwarzschild radius, $r_s$. 
The horizon becomes prolate as the value $b/a$ increases. 
We can not find the apparent horizon when length $b$ is larger than $b=1.5$ for $a=0.5$ and $b=2.0$ for $a=0.1$.
We see from (b) and (e) of Fig.2 that the matter configurations can be arbitrarily large but the apparent horizon does not cover all the matter regions.
This behavior is the same with $3+1$ dimensional cases \cite{nakamura} and our numerical results reproduce the results in \cite{yoo}.
If we compare our 5-dimensional results with 4-dimensional ones (ref.[4]), the disappearance of the apparent horizon 
can be seen only for the highly prolate cases. (E.g., for the eccentricity 0.999 cases, the disappearance of the apparent horizon
starts at the prolate radius 0.7 $M$ in 4-dimensional case, while 2.0 $r_s$ in our case.)  Therefore we expect that an appearance 
of a singular behaviour is ``relaxed" in 5-dimensional case, and this tendency would be the same for the higher-dimensional cases. 
\begin{figure*}[htbp] 
  \begin{center}
    \begin{tabular}{ccc}
	  \resizebox{50mm}{!}{\includegraphics{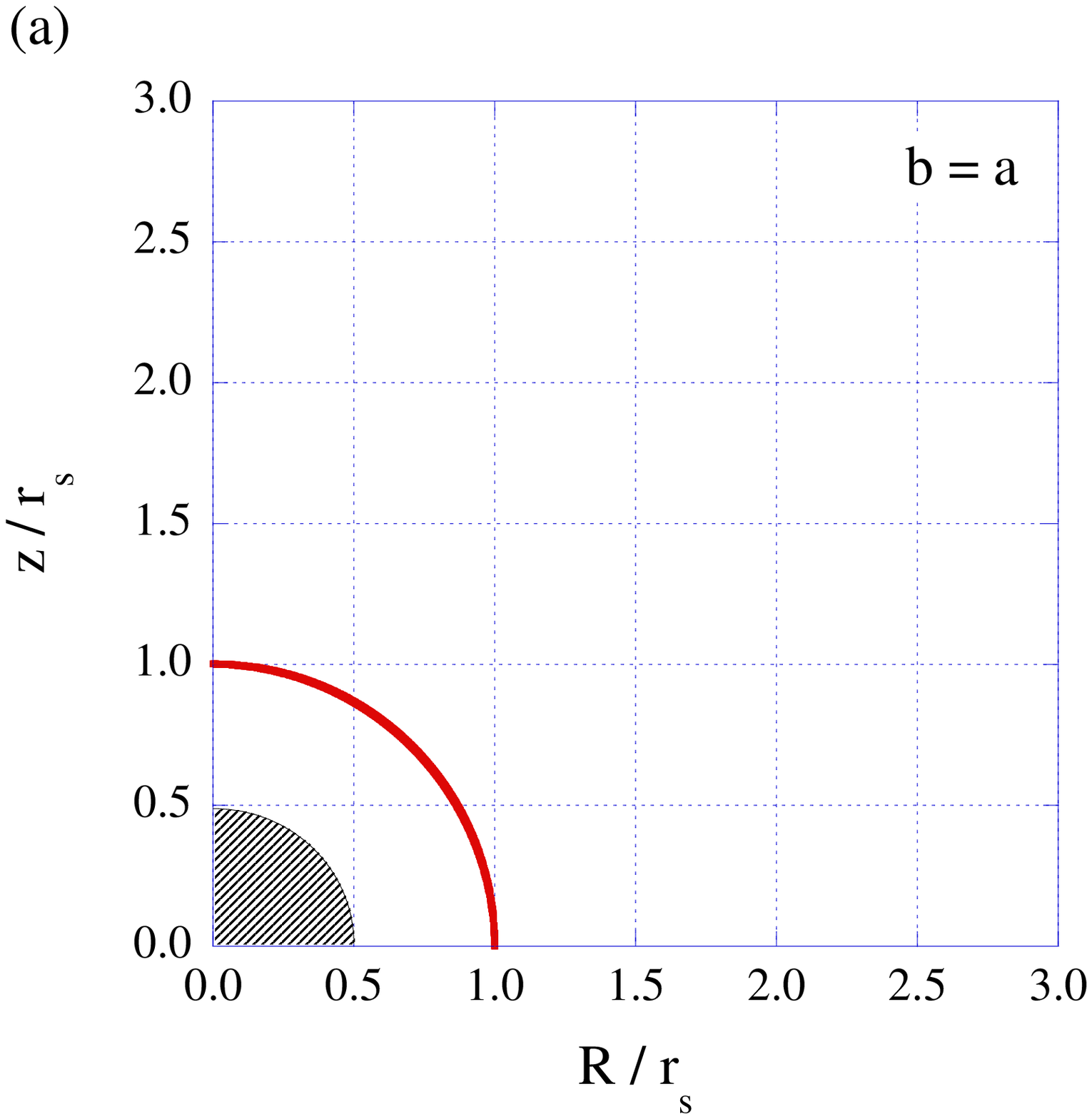}} &
      \resizebox{50mm}{!}{\includegraphics{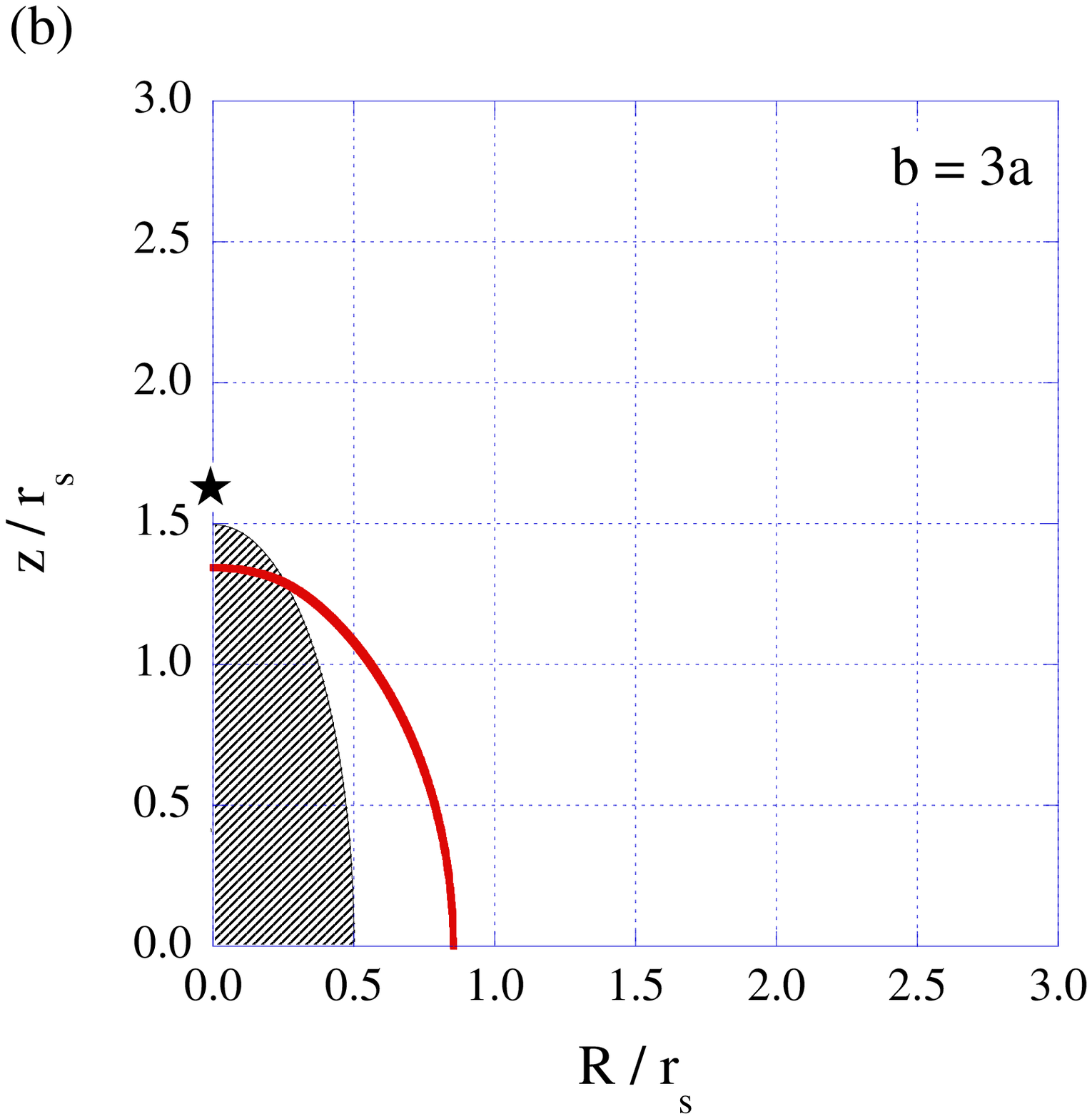}} &
      \resizebox{50mm}{!}{\includegraphics{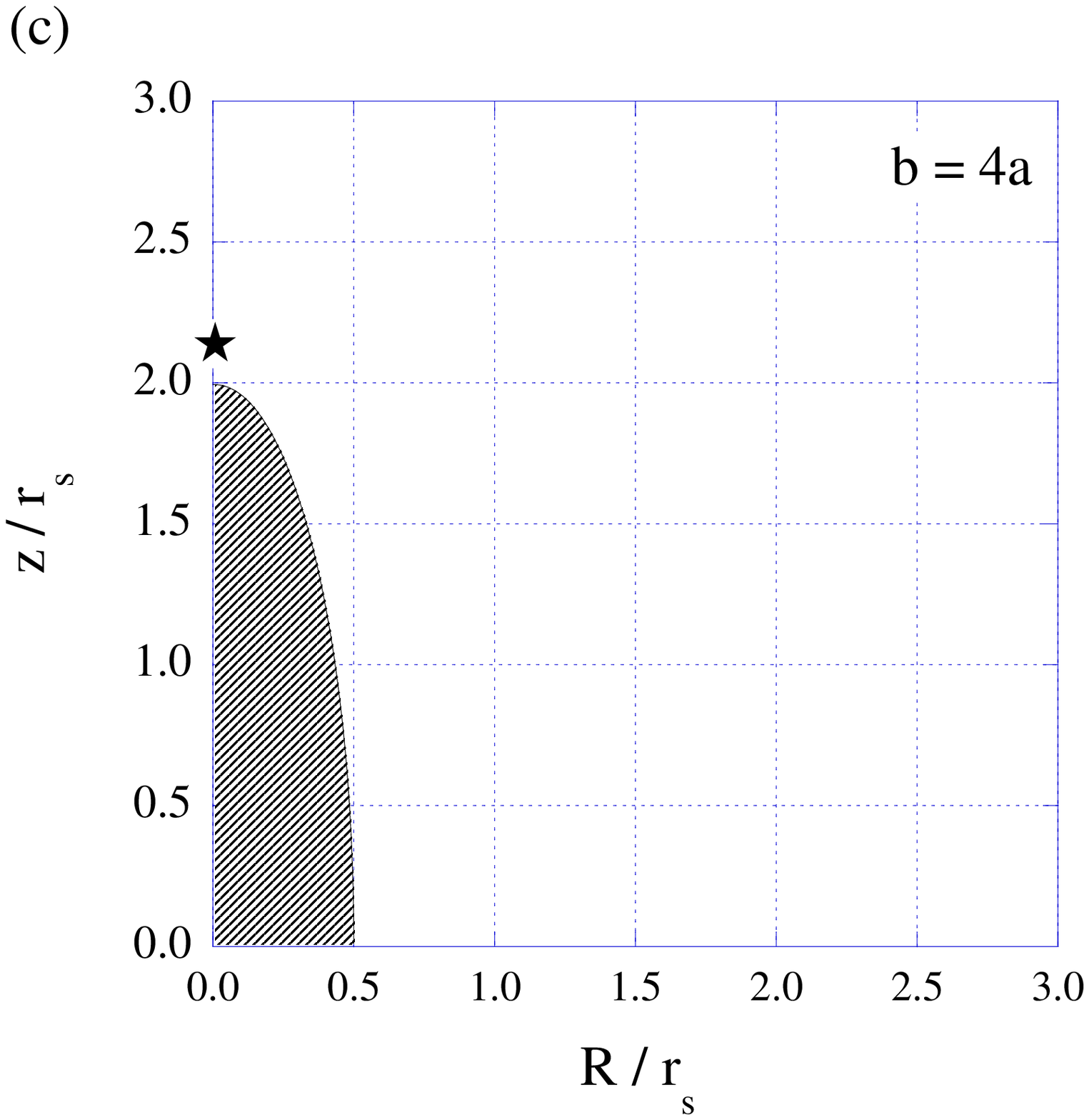}}\\
    \end{tabular}
  \end{center}
  \begin{center}
    \begin{tabular}{ccc}
      \resizebox{50mm}{!}{\includegraphics{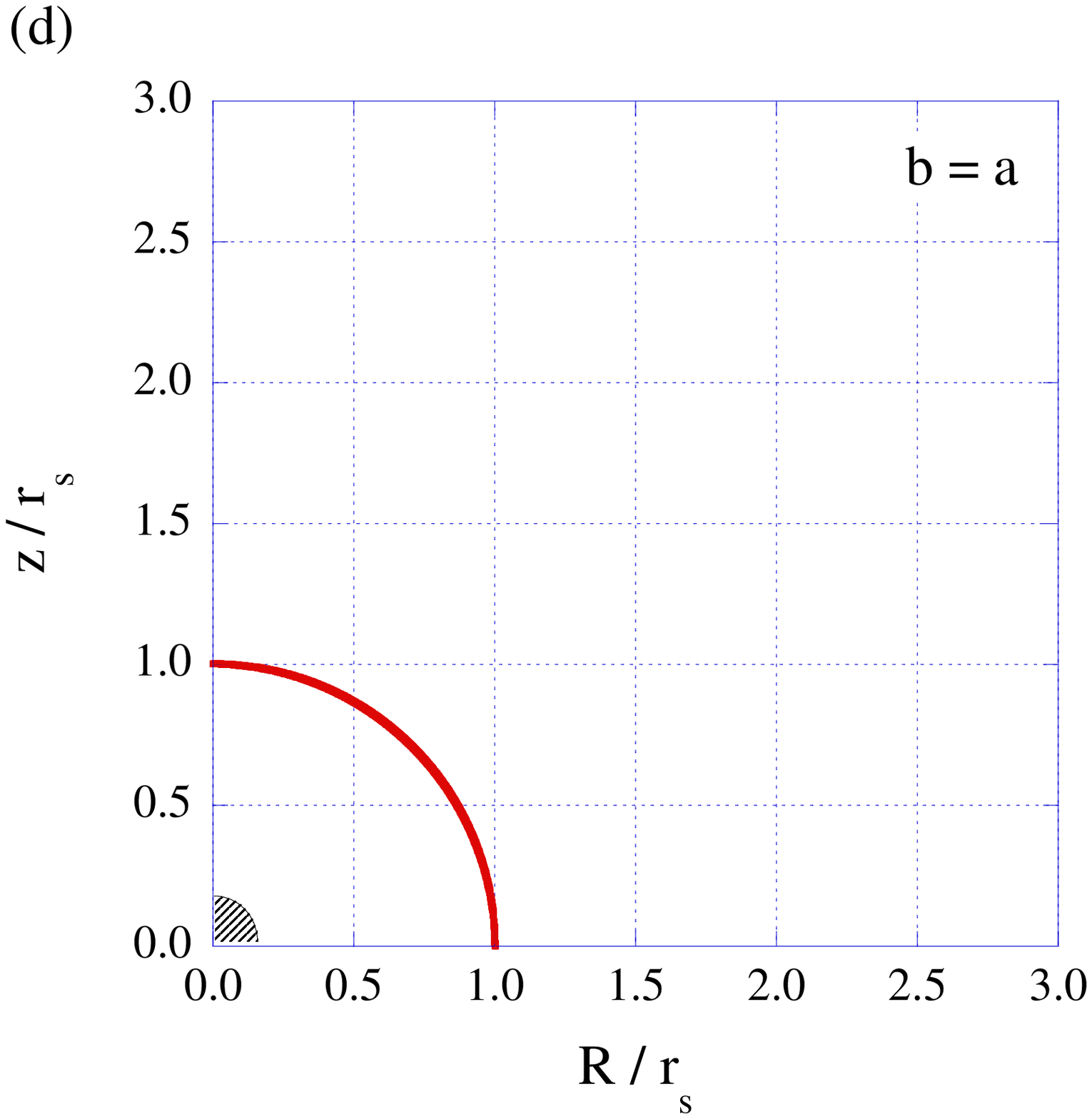}} &
      \resizebox{50mm}{!}{\includegraphics{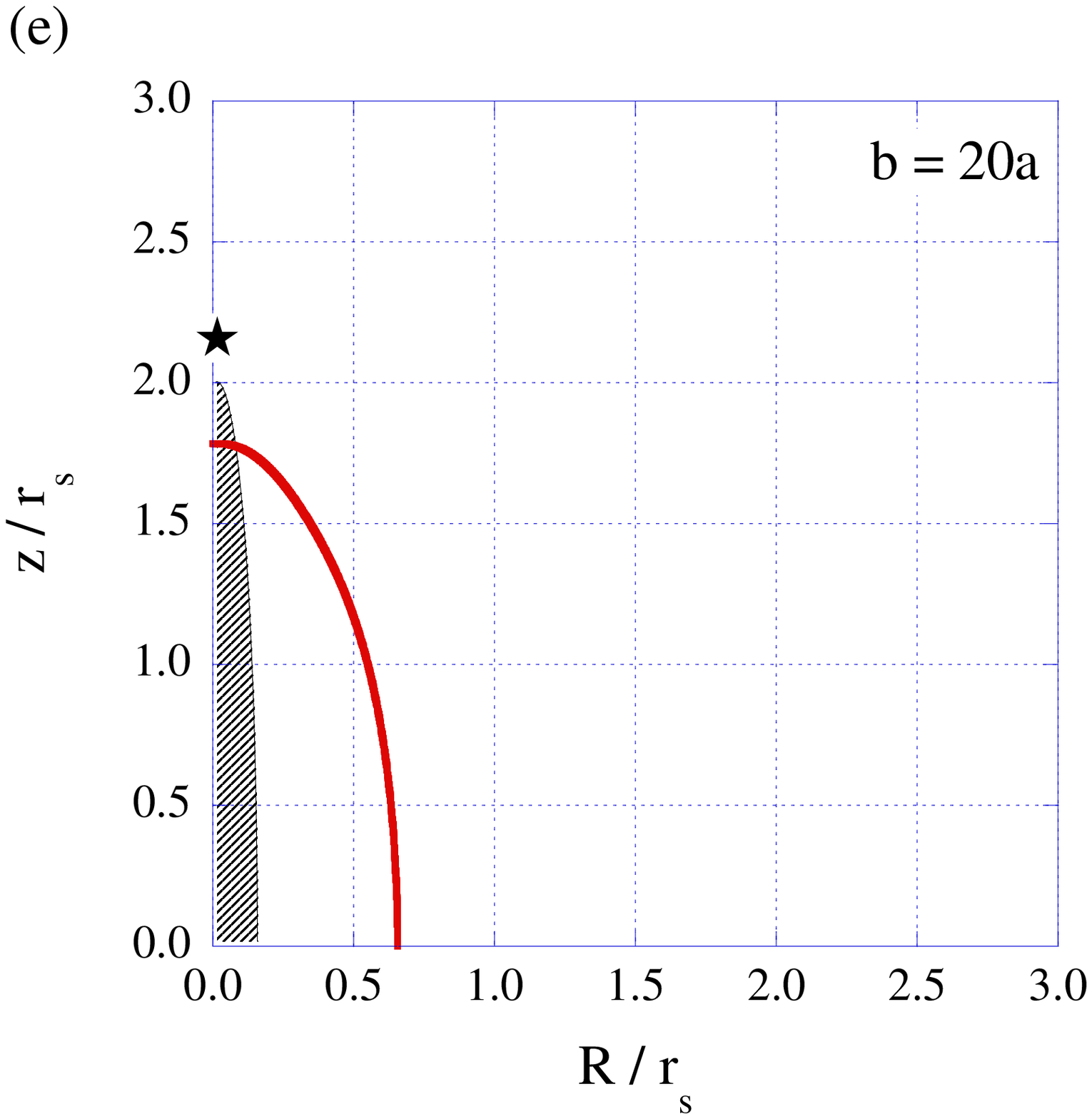}} &
      \resizebox{50mm}{!}{\includegraphics{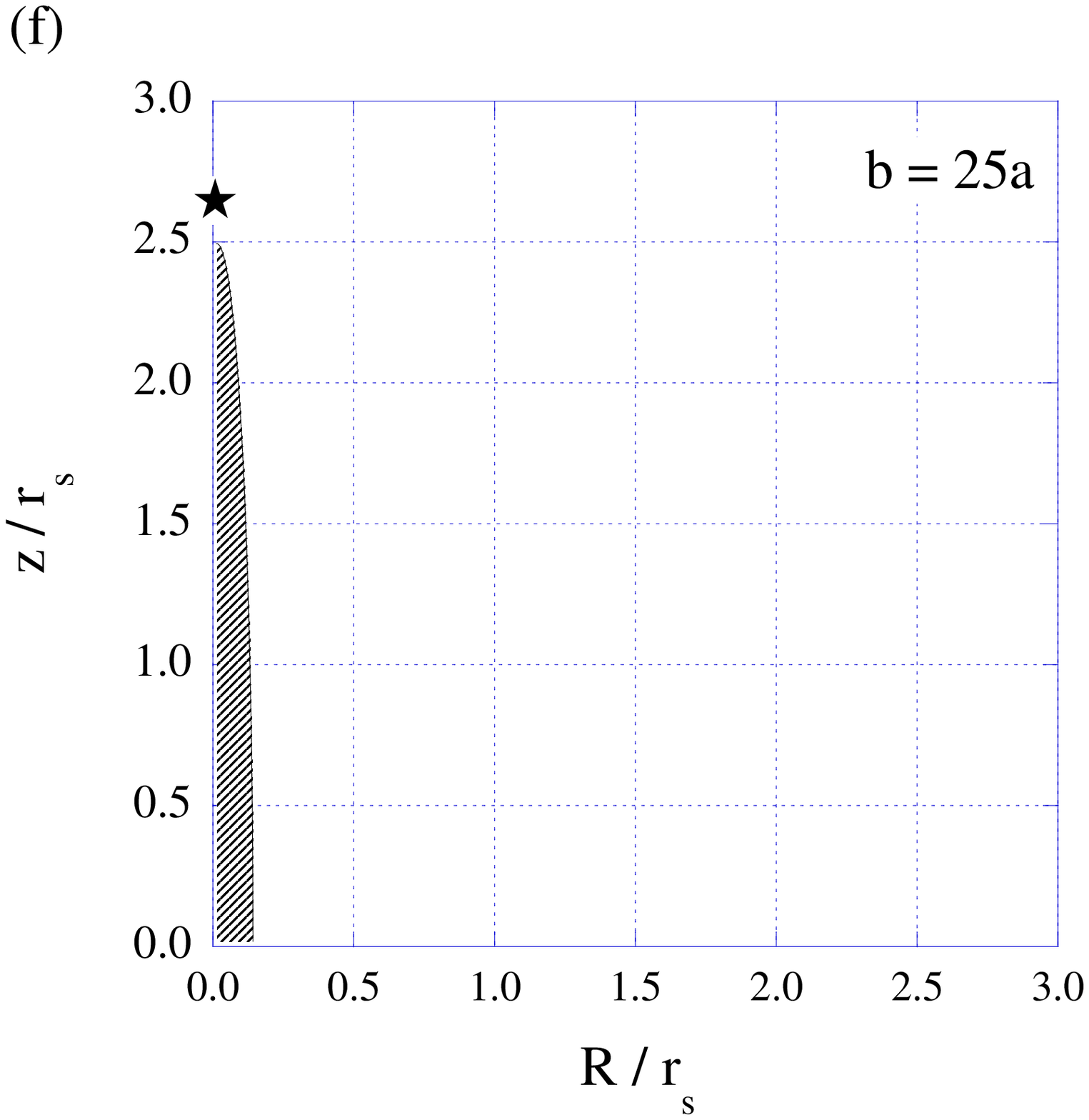}}\\
    \end{tabular}
    \caption{\label{fig1}Matter distributions (shadows) and apparent horizons (lines) for spheroidal matter distributions. The sections of axis-equator plane are shown. The sequence (a)-(c) is of $a=0.5$, and (d)-(f) is of $a=0.1$
$\left[\right.$see Eq.(\ref{spindle})$\left.\right]$, of which we fix the total mass $M_{ADM}=1$. 
We can not find an apparent horizon when $b$ is larger than $b=3a$ for $a=0.5$$\left[\right.$Fig.(c)$\left.\right]$ and $b=20a$ for $a=0.1$$\left[\right.$Fig.(f)$\left.\right]$. The asterisks indicate the location of the maximum Kretchmann invariant, Eq.(\ref{invariant}). We see the maximum point is outside of the horizon for the case (b) and (e).}
  \end{center}
\end{figure*}
\begin{figure*}[htbp] 
  \begin{center}
    \begin{tabular}{ccc}
	  \resizebox{50mm}{!}{\includegraphics{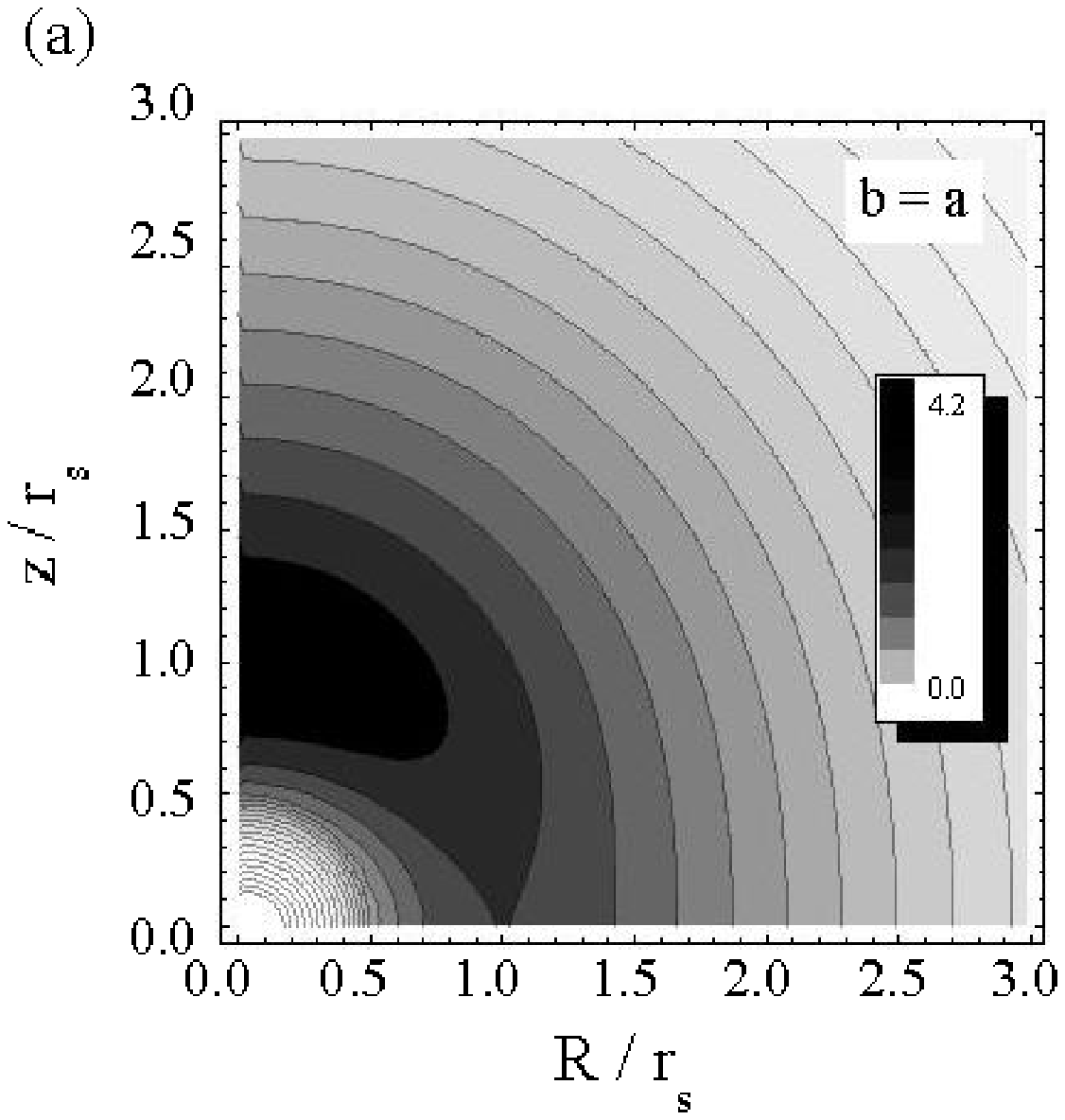}} &
      \resizebox{50mm}{!}{\includegraphics{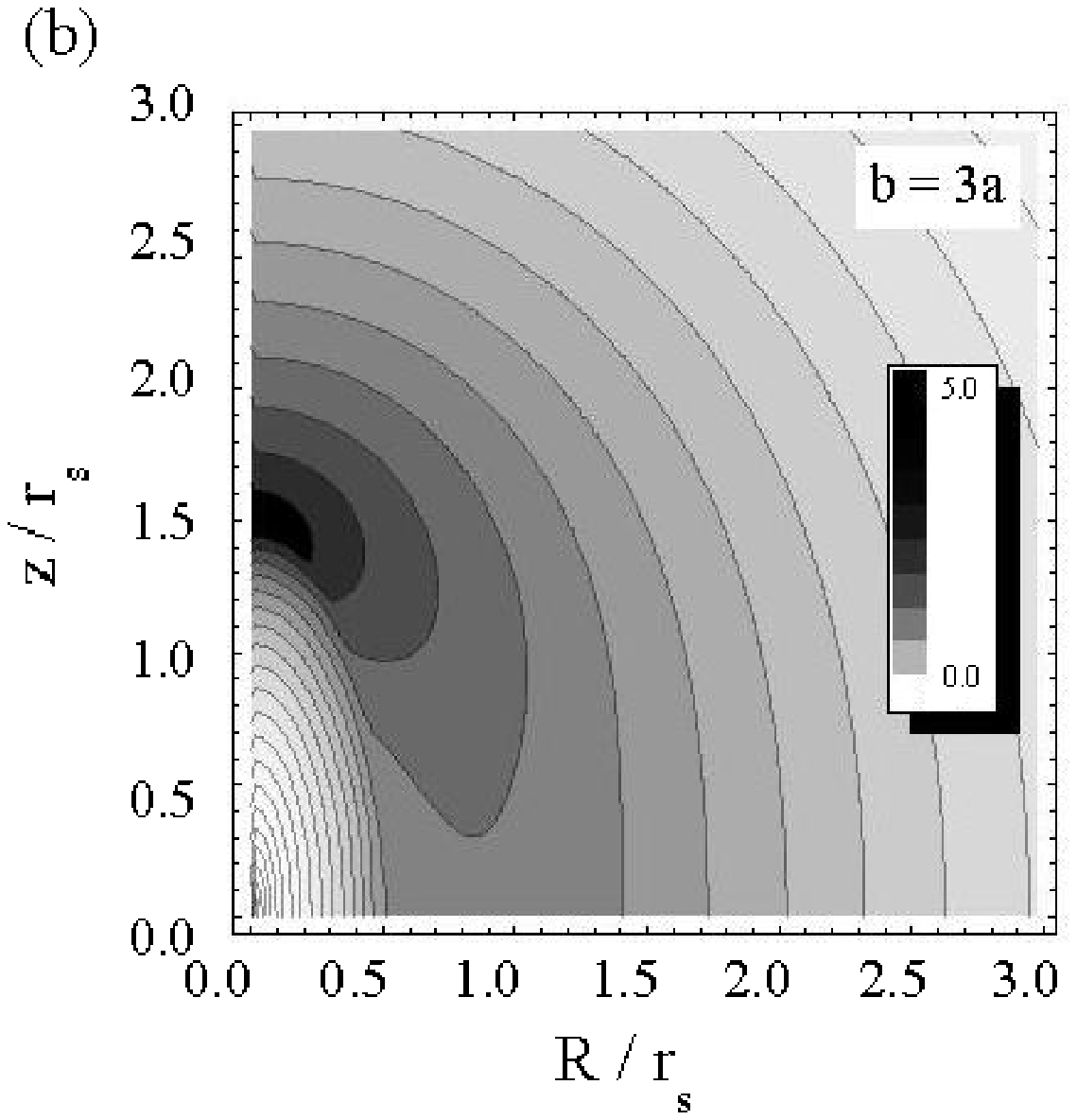}} &
      \resizebox{50mm}{!}{\includegraphics{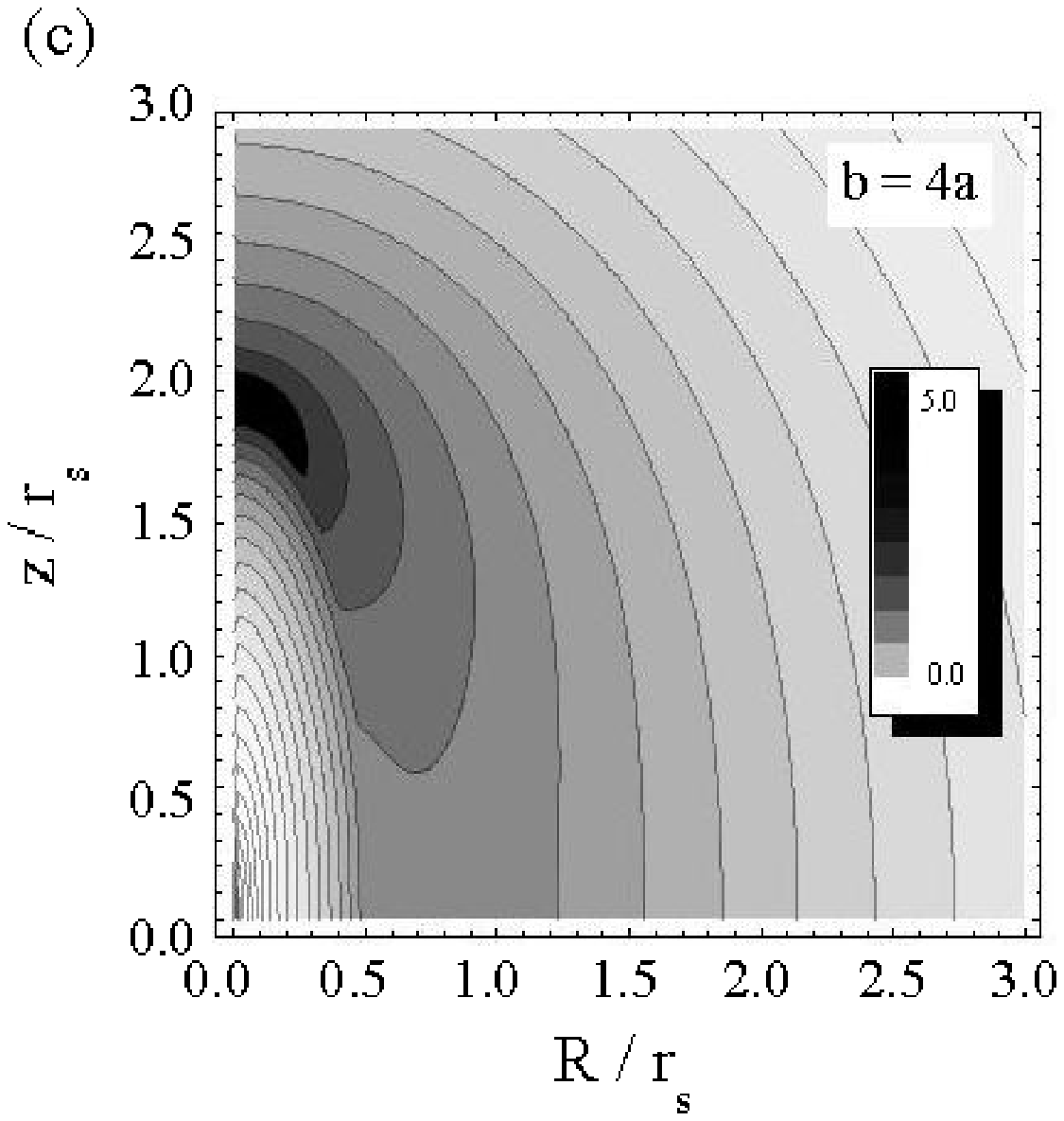}}\\
    \end{tabular}
  \end{center}
  \begin{center}
    \begin{tabular}{ccc}
      \resizebox{50mm}{!}{\includegraphics{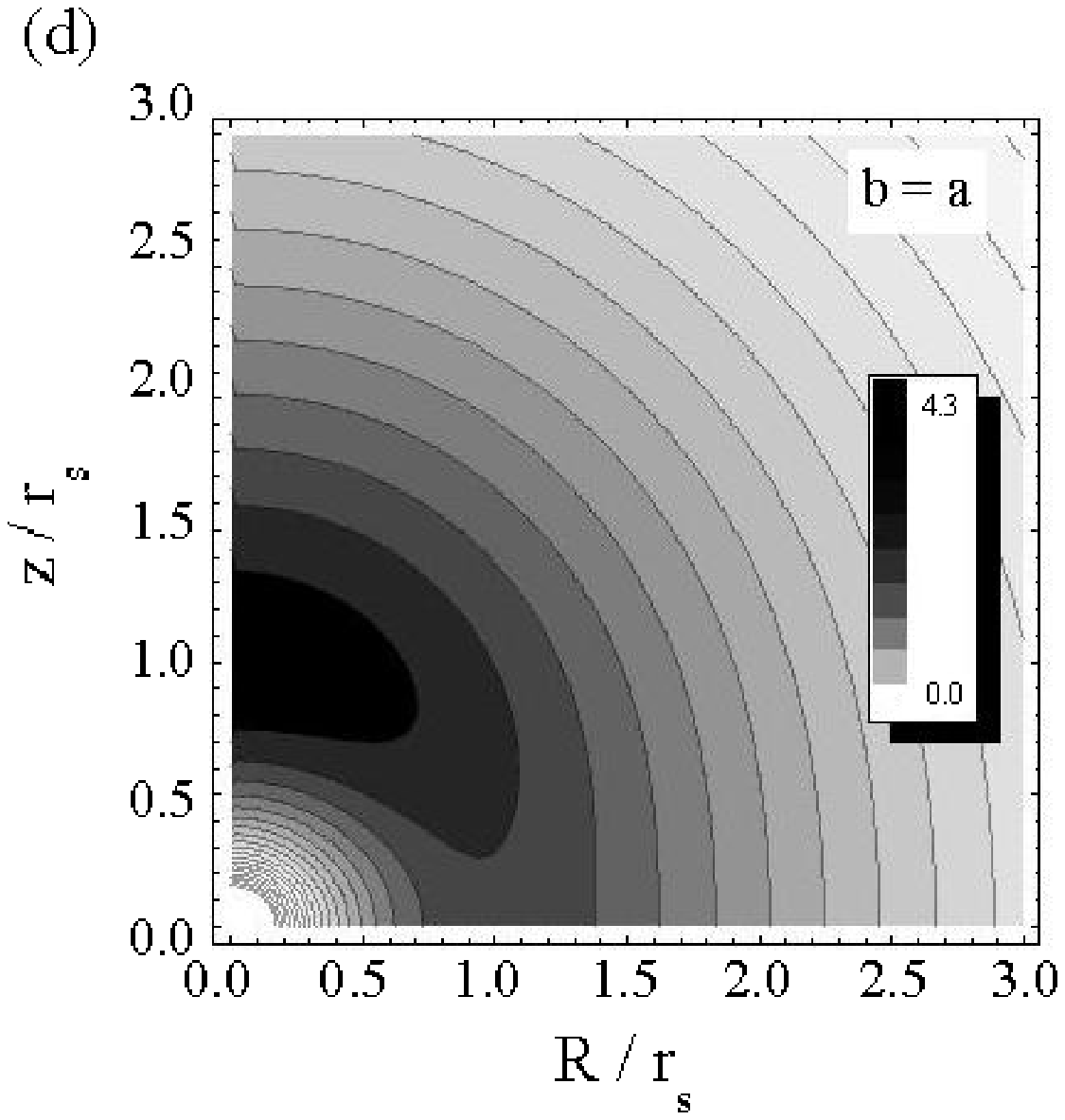}} &
      \resizebox{50mm}{!}{\includegraphics{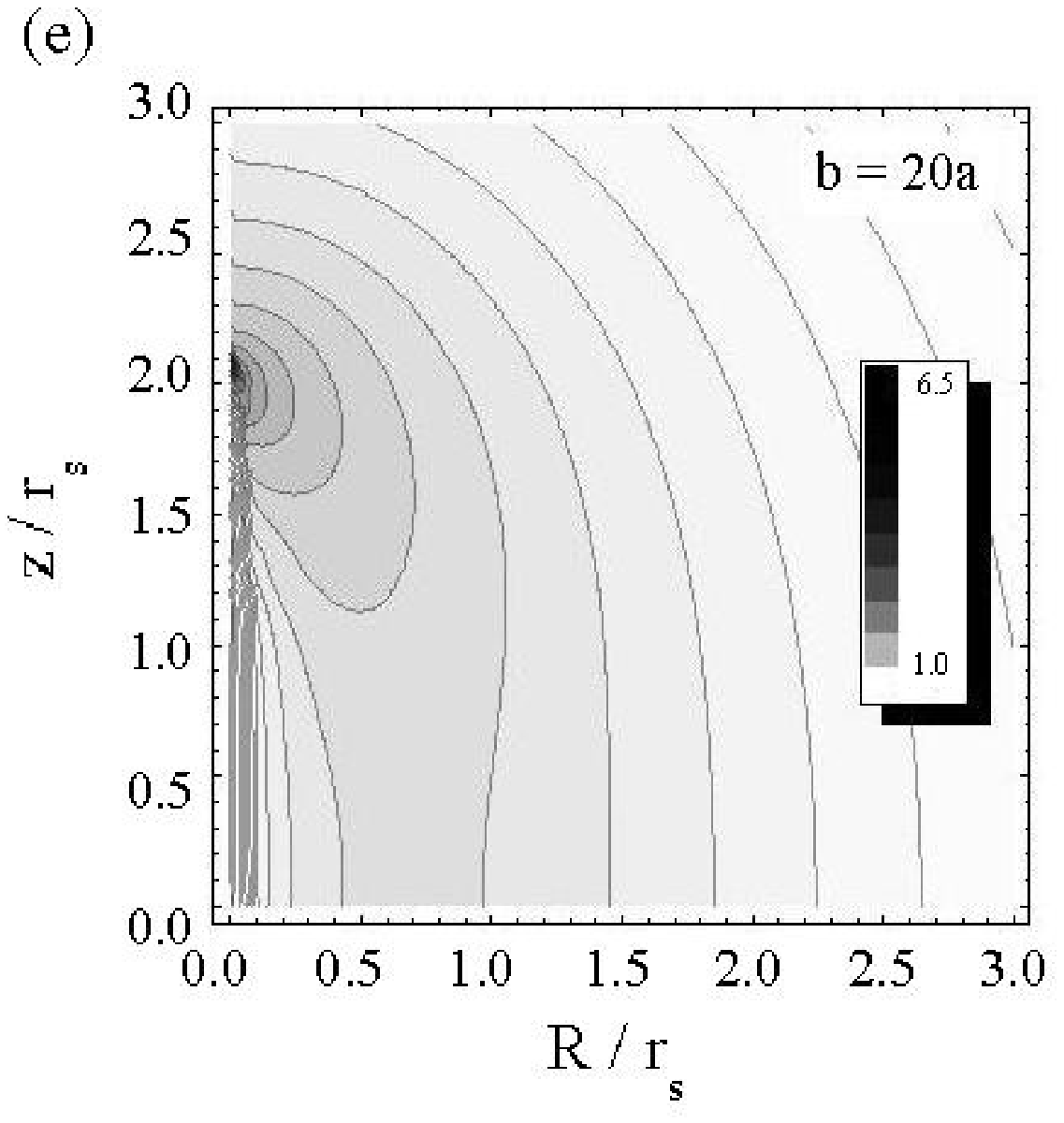}} &
      \resizebox{50mm}{!}{\includegraphics{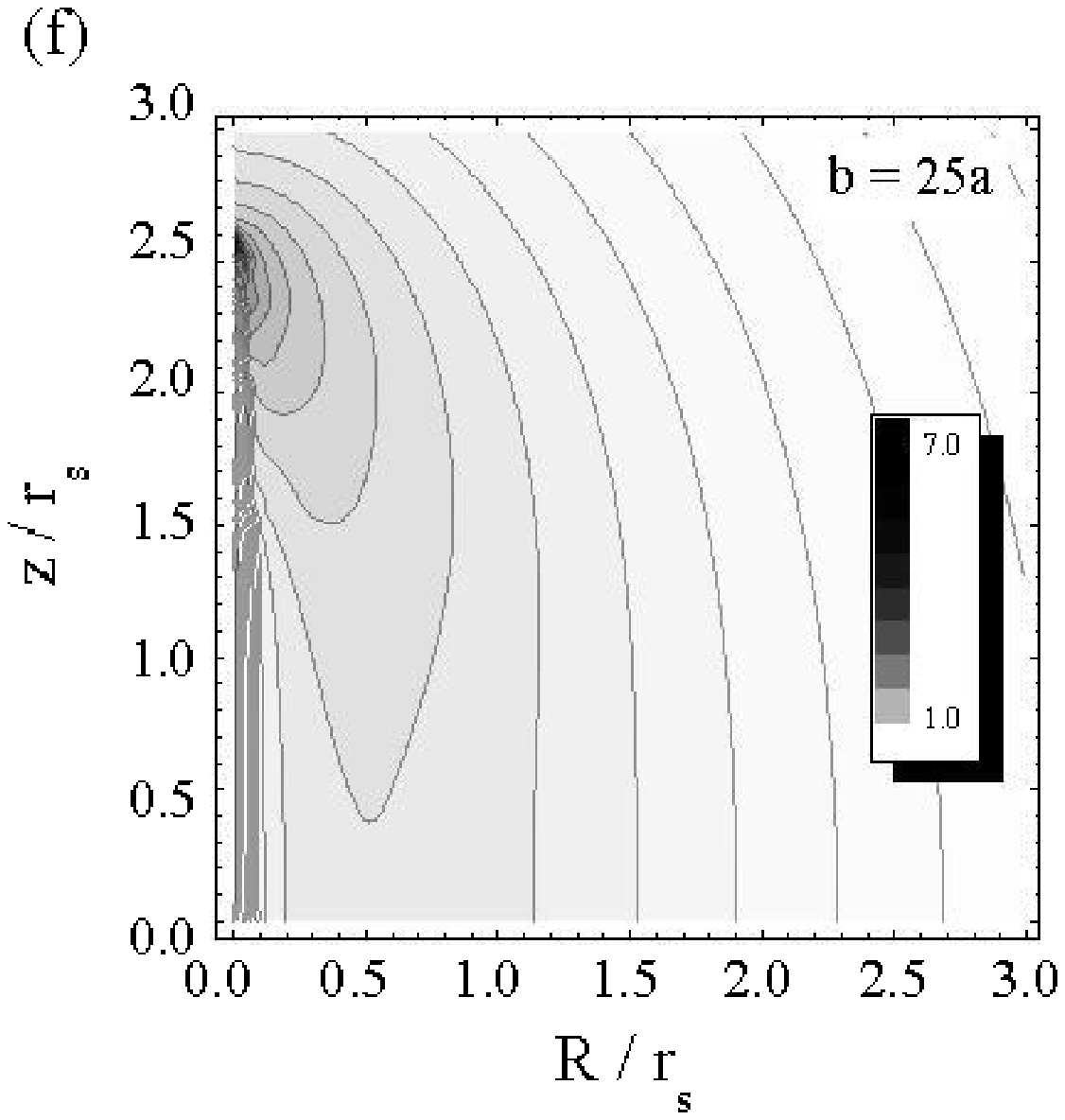}}\\
    \end{tabular}
    \caption{\label{C}Contours of Kretchmann invariant, $\log_{10}{\cal I}^{(4)}$, corresponding to Fig.\ref{fig1}.}
  \end{center}
\end{figure*}

\begin{figure*}[htbp] 
  \begin{center}
    \begin{tabular}{cc}
      \resizebox{80mm}{!}{\includegraphics{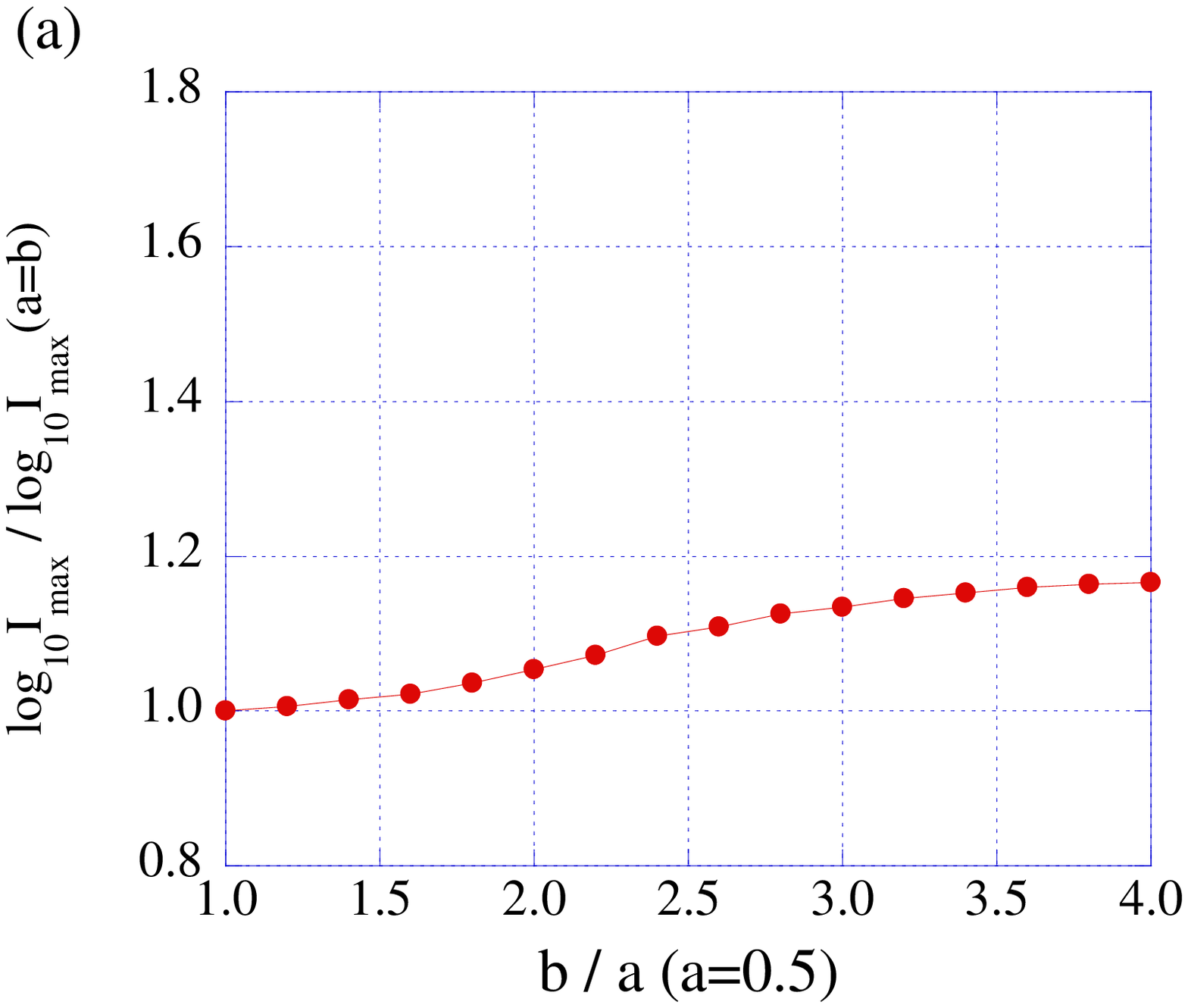}} &
      \resizebox{80mm}{!}{\includegraphics{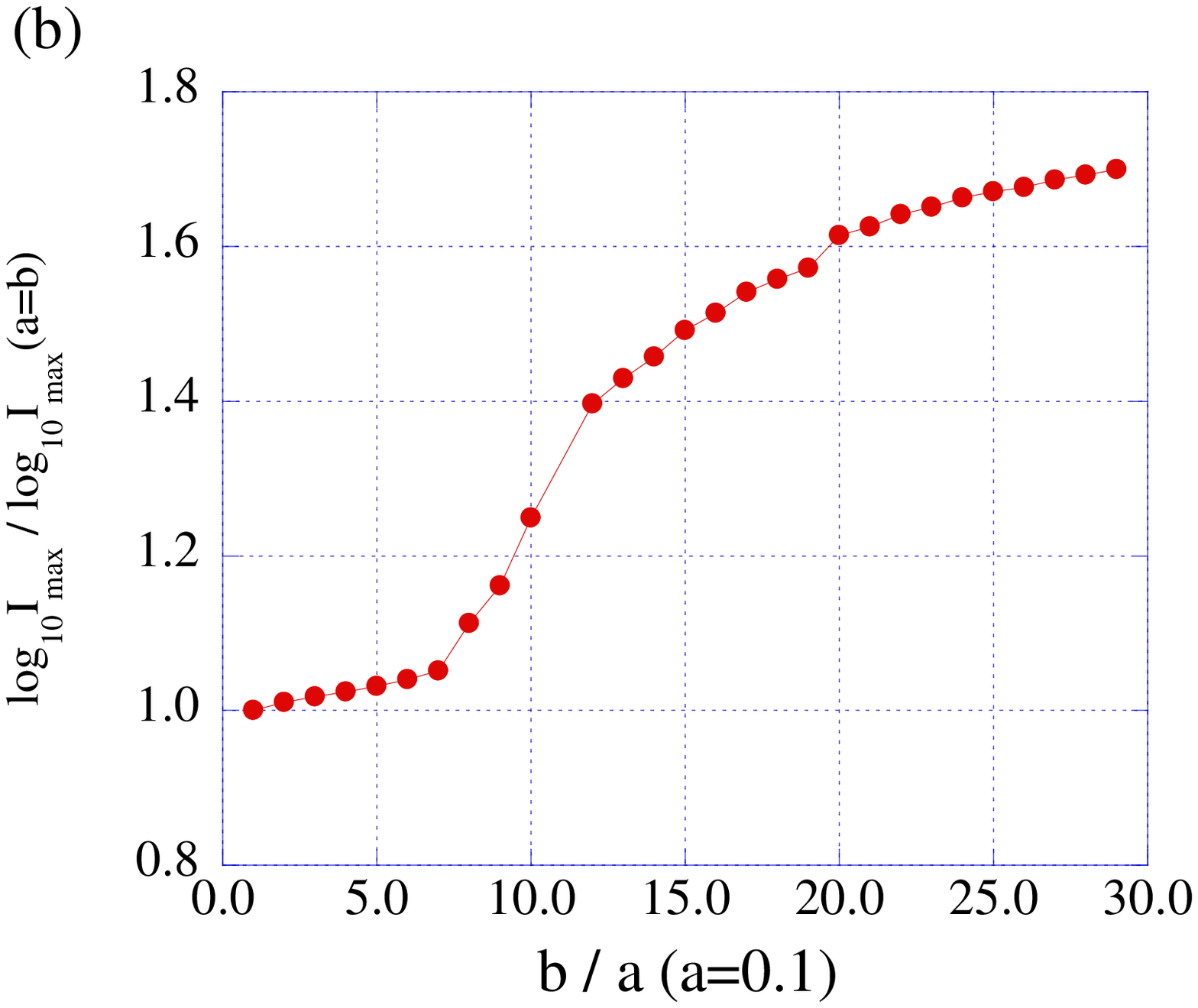}}
    \end{tabular}
    \caption{\label{Imax_spindle}The maximum value of Kretchmann invariant ${\cal I}_{max}$ as a function of $b/a$ for the sequences of Fig.2. Plots are normalized with the value of the spherical case, $a=b$. We see that ${\cal I}_{max}$ increases monotonically both cases.}
  \end{center}
\end{figure*}

The asterisk in Fig.2 is the location of the largest Kretchmann invariant, ${\cal I}_{\rm max}={\rm max} \{ R_{abcd}^{(4)}R^{(4)abcd}\}$.
For all cases, we see the locations of ${\cal I}_{\rm max}$ are always outside the matter, except the cases of $b=a$.\footnote{The Kretchmann invariant expresses the strength of the curvature, which is determined by the gradient of metric. For example, when we solve a single star with uniform density, the maximum value of the metric appears at the center of matter configuration, but the maximum value of the metric gradient appears off-center and likely at the outside of matter region. Therefore, our results of the location of the maximum Kretchmann invariant is not strange.}
We show the contours of ${\cal I}^{(4)}$ in Fig.\ref{C}.
Fig.\ref{Imax_spindle} displays ${\cal I}_{max}$ as a function of $b/a$.
We see that ${\cal I}_{max}$ monotonically increases even if there is no apparent horizon.
In the $3+1$ dimensional cases, the extremely elongated spindle evolves into a naked singularity \cite{nakamura}.
Our results suggest such evolutions also in the $4+1$ dimensional cases.

\begin{figure*}[htbp] 
  \begin{center}
    \begin{tabular}{cc}
      \resizebox{80mm}{!}{\includegraphics{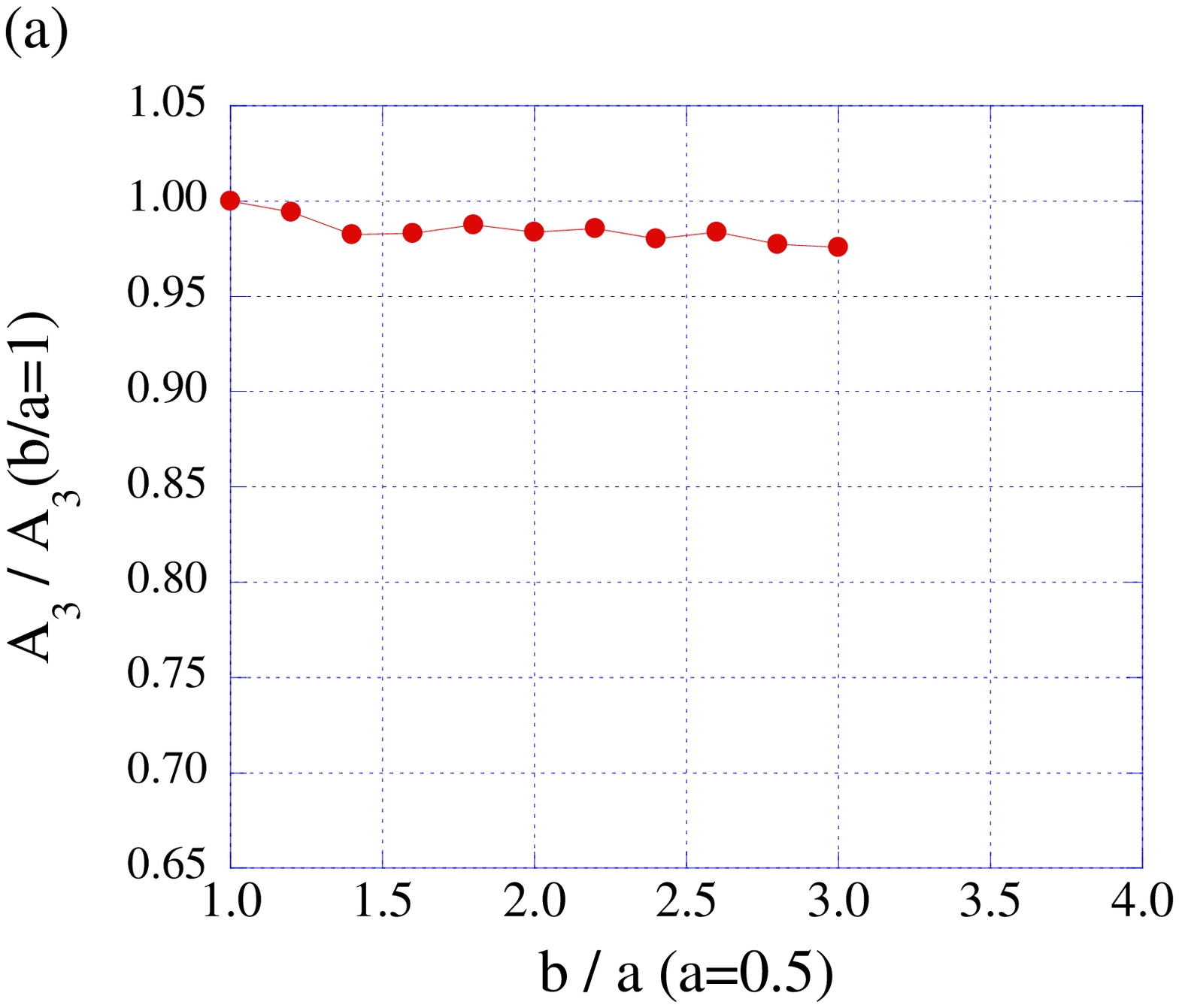}} &
      \resizebox{80mm}{!}{\includegraphics{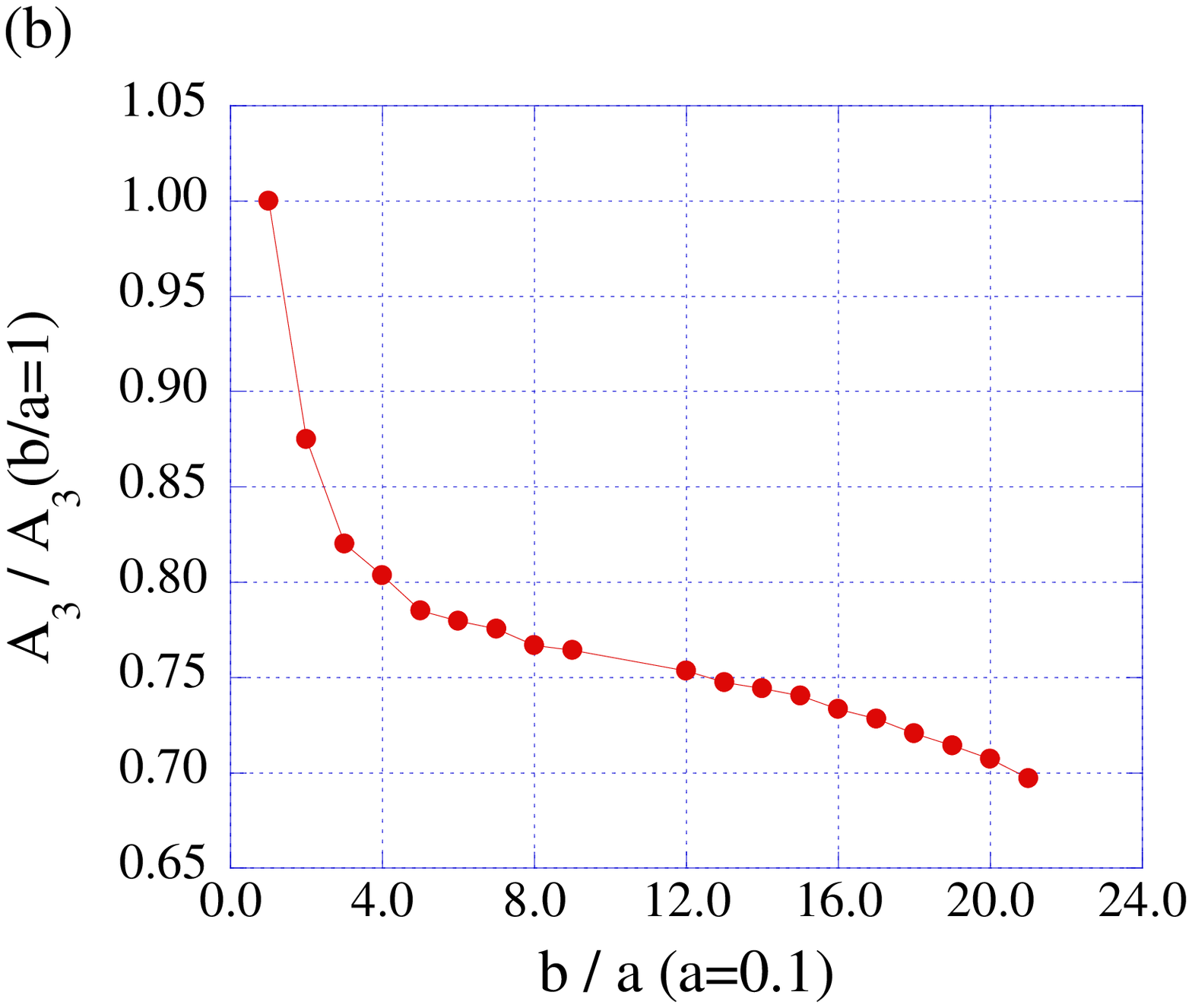}}
    \end{tabular}
    \caption{\label{A}The area of the apparent horizon $A_3$ for the sequence of Fig.2 is shown. The sequence of $a=0.5$ and $0.1$ is shown in (a) and (b), respectively. Plots are normalized with the area of the  spherical case, $a=b$. In both cases, the horizon area monotonically decreases with $b/a$.}
  \end{center}
\end{figure*}

In Fig.\ref{A}, we show the surface area of the apparent horizon $A_3$.
We observe $A_3$ becomes the largest when the matter is spherical.
If we took account the analogy of the thermodynamics of black-hole, this may  suggest that the final state of 5D black-hole shakes down to spherically symmetric.

\begin{figure*}[htbp] 
  \begin{center}
    \begin{tabular}{cc}
      \resizebox{80mm}{!}{\includegraphics{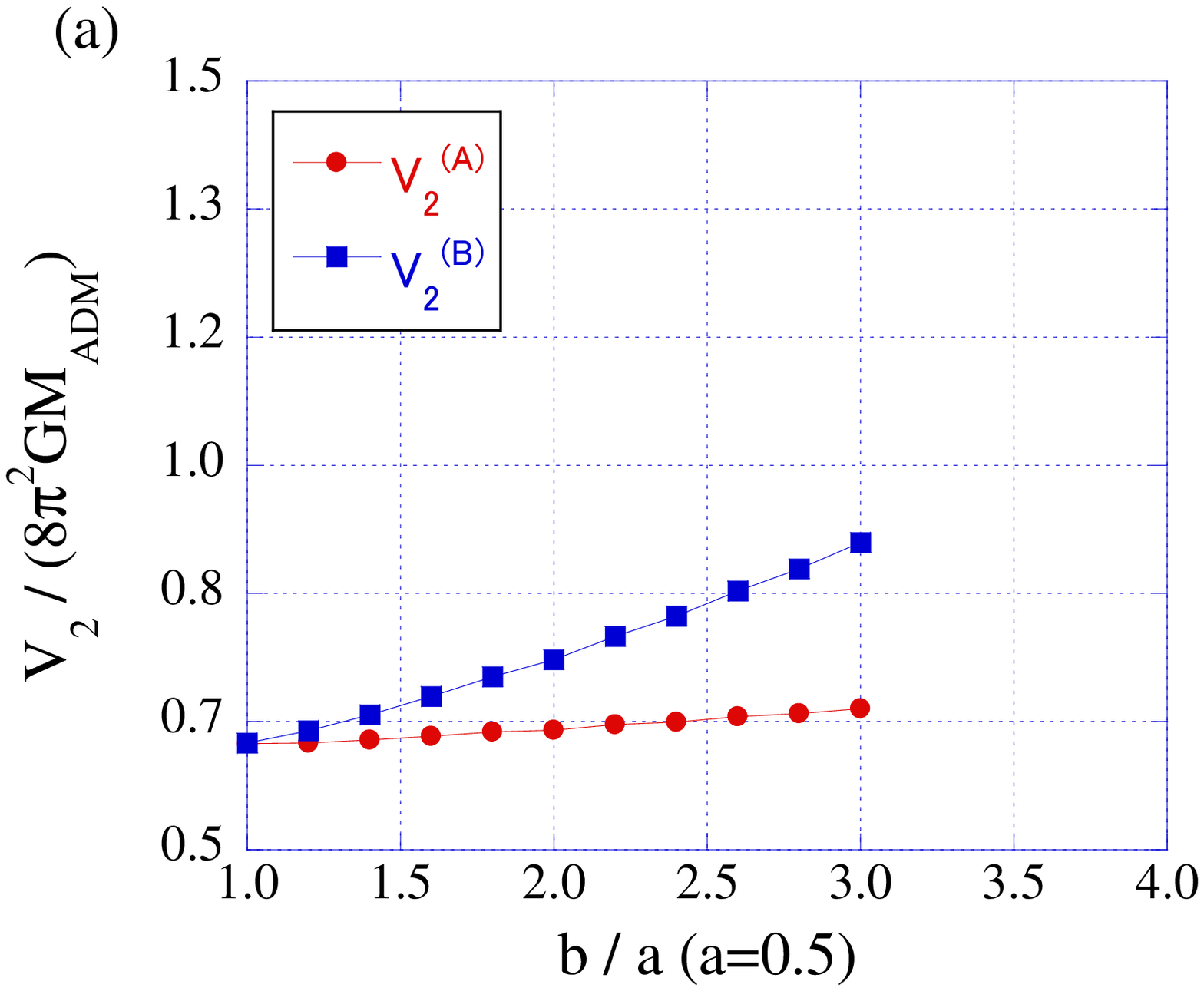}} &
      \resizebox{80mm}{!}{\includegraphics{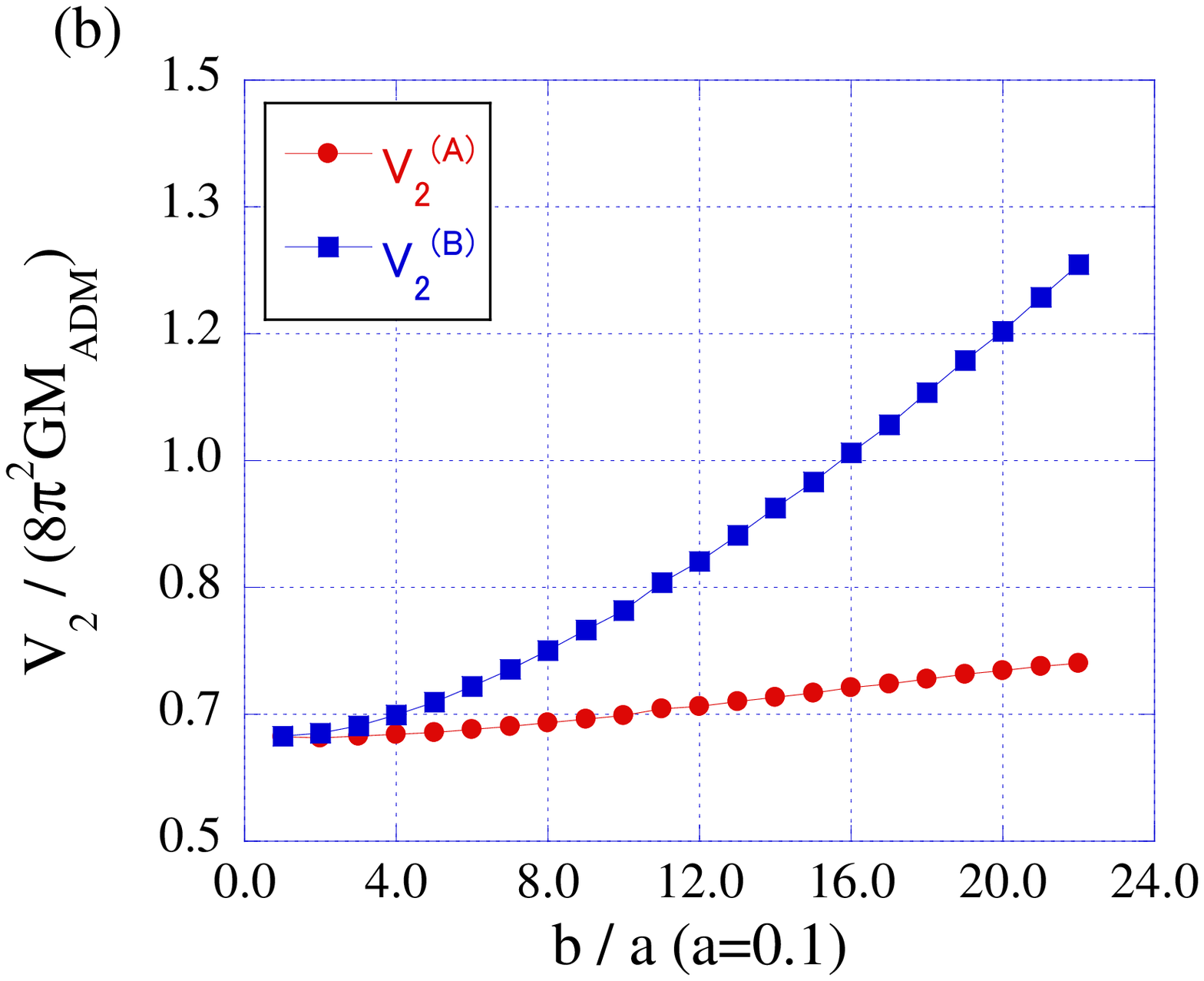}}
    \end{tabular}
    \caption{\label{V2}The ratio of the hyper-hoop $V_2$ to the mass $M_{ADM}$ is shown for the sequence of Fig.2. The ratio less than unity indicates that the validity of the hyper-hoop conjecture, Eq.(\ref{sufficient_condition}). We plot the hoops $V_2^{(A)}$ and $V_2^{(B)}$ both for the sequences of $a=0.5$ and $0.1$ in Fig.(a) and (b), respectively. At large $b/a$, the hoops do not exist, but that range always includes the cases with apparent-horizon formation. Figure shows that the hoop $V_2^{(A)}$ represents the hyper-hoop conjecture properly.}
  \end{center}
\end{figure*}

In order to check the validity of the hyper-hoop conjecture, we prepared Fig.\ref{V2}.
The hyper-hoop $V_2^{(A)}$ and $V_2^{(B)}$ are shown with the normalized value with the right-hand side of  Eq.(\ref{sufficient_condition}), i.e. the validity of the conjecture indicates the value is less than unity.
The area of the hyper-hoops $V_2^{(A)}$ and $V_2^{(B)}$ increase with $b/a$ but $V_2^{(A)}$ remains smaller than unity if the horizon exists.
Therefore the necessary condition of black-hole formation $\left[\right.$Eq.(\ref{sufficient_condition})$\left.\right]$ is satisfied for $V_2^{(A)}$.
We conclude that hyper-hoop conjecture is valid for the spheroidal cases.
\clearpage
\subsection{Toroidal configurations}
\begin{figure}[htbp] 
  \begin{center}
    \begin{tabular}{ccc}
      \resizebox{50mm}{!}{\includegraphics{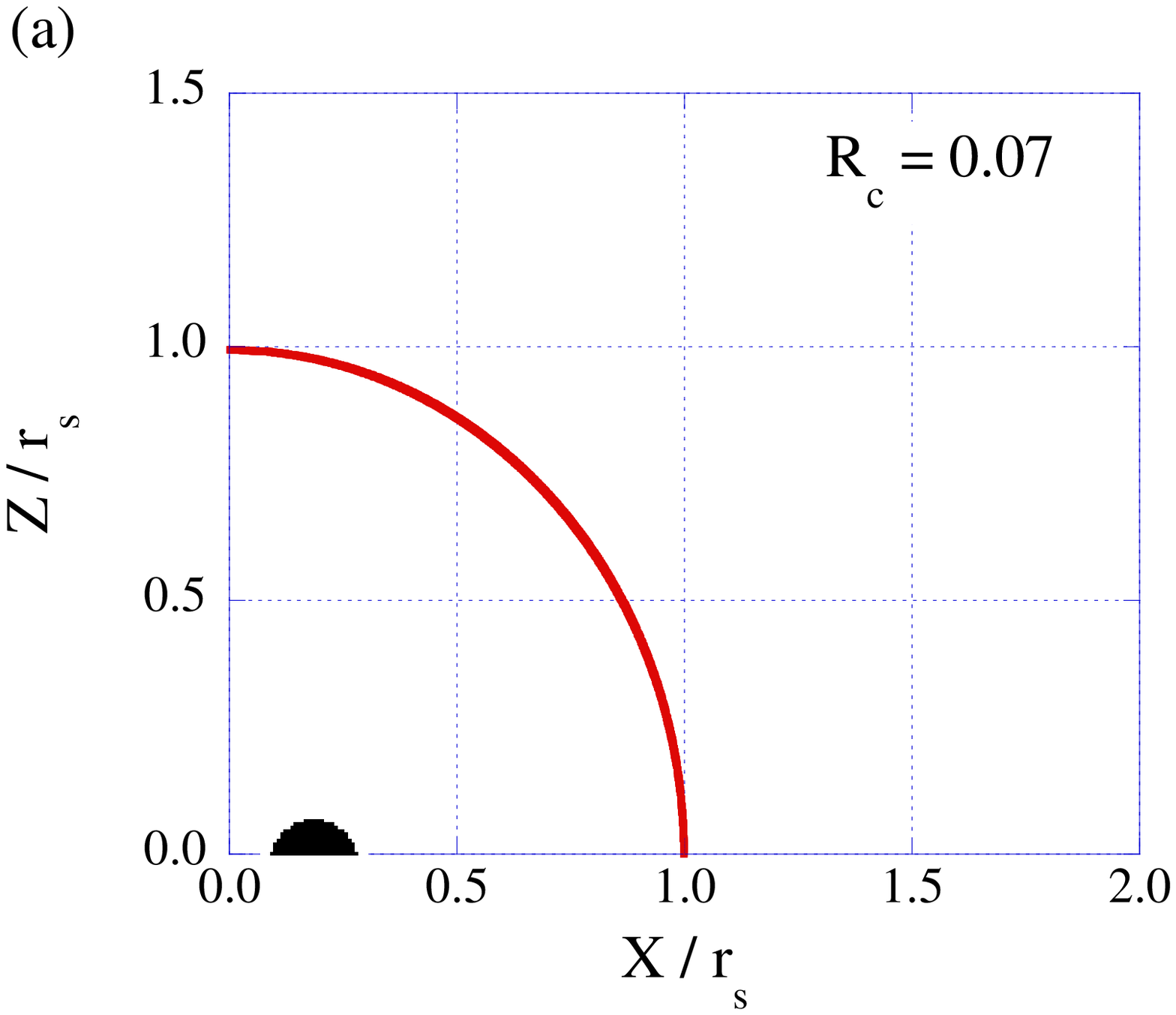}} &
      \resizebox{50mm}{!}{\includegraphics{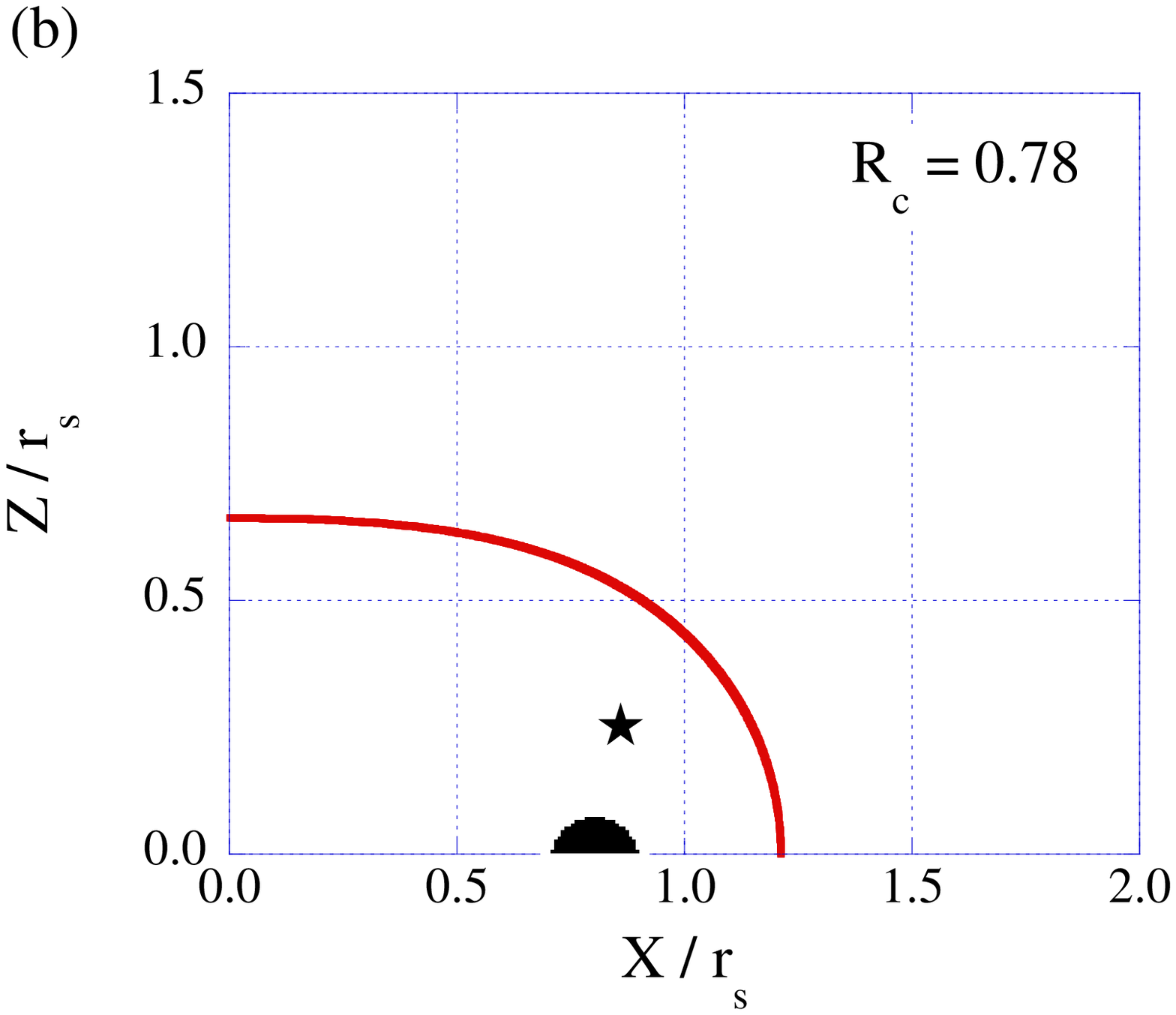}} &
      \resizebox{50mm}{!}{\includegraphics{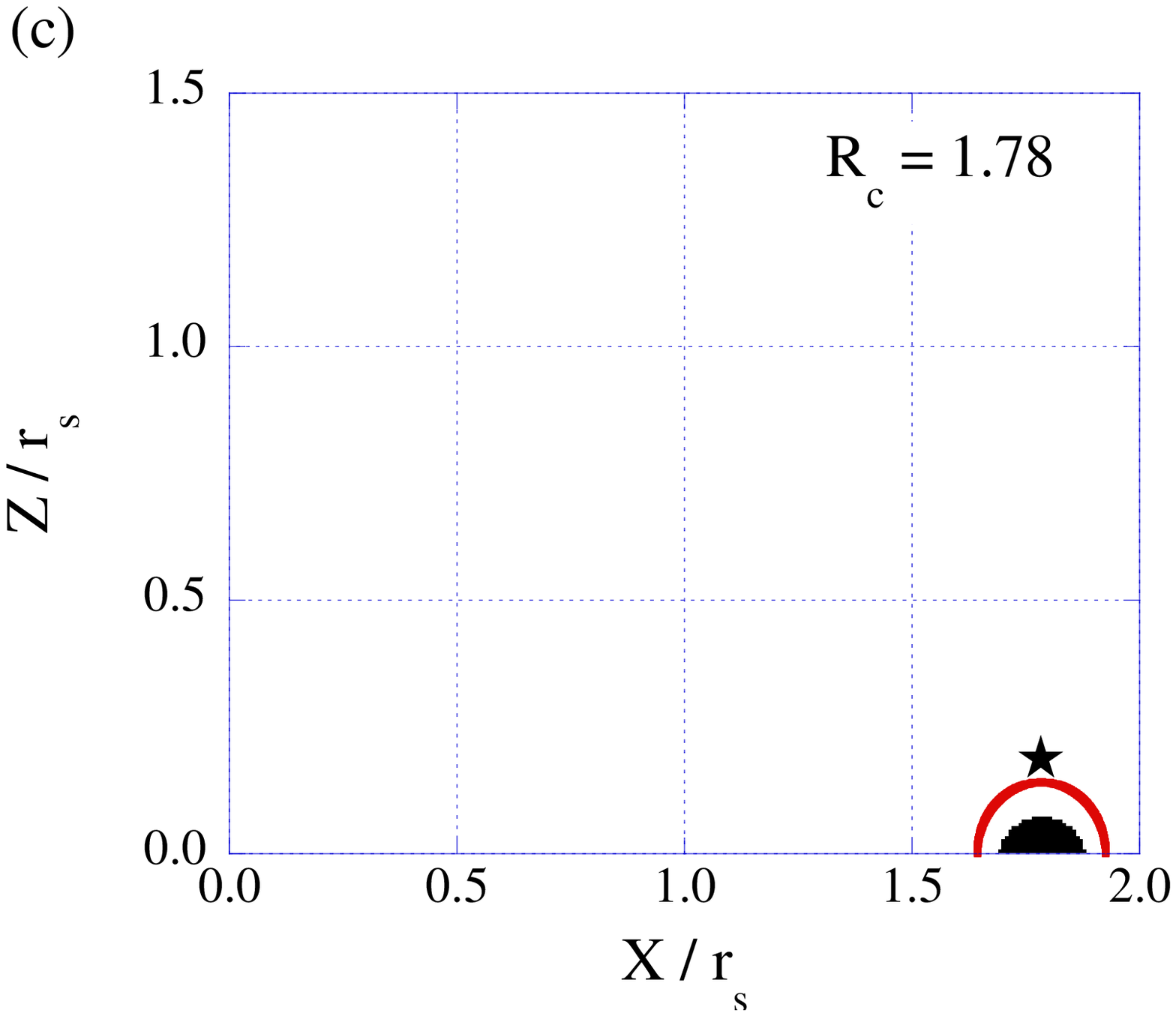}}\\
    \end{tabular}
    \caption{\label{horizon_r0.1}Matter distributions (shaded) and the location of the apparent horizon (line) for toroidal matter configurations (with fixing the ring radius $R_r=0.1$). The axis-equator plane is shown for three circle-radius cases:(a) $R_c=0.07$, (b) $R_c=0.78$, and (c) $R_c=1.78$ $\left[\right.$see Eq.(\ref{toroidal})$\left.\right]$. Line is the location of the apparent horizon. We found the common horizon ($S^3$) for (a) and (b), while we found the ring horizon ($S^1 \times S^2$) for (c). The asterisk indicates the location of the maximum Kretchmann invariant, ${\cal I}_{max}$. We see the maximum point is outside of the horizon for the case (c).}
  \end{center}
\end{figure}

\begin{figure}[htbp] 
  \begin{center}
    \begin{tabular}{ccc}
      \resizebox{50mm}{!}{\includegraphics{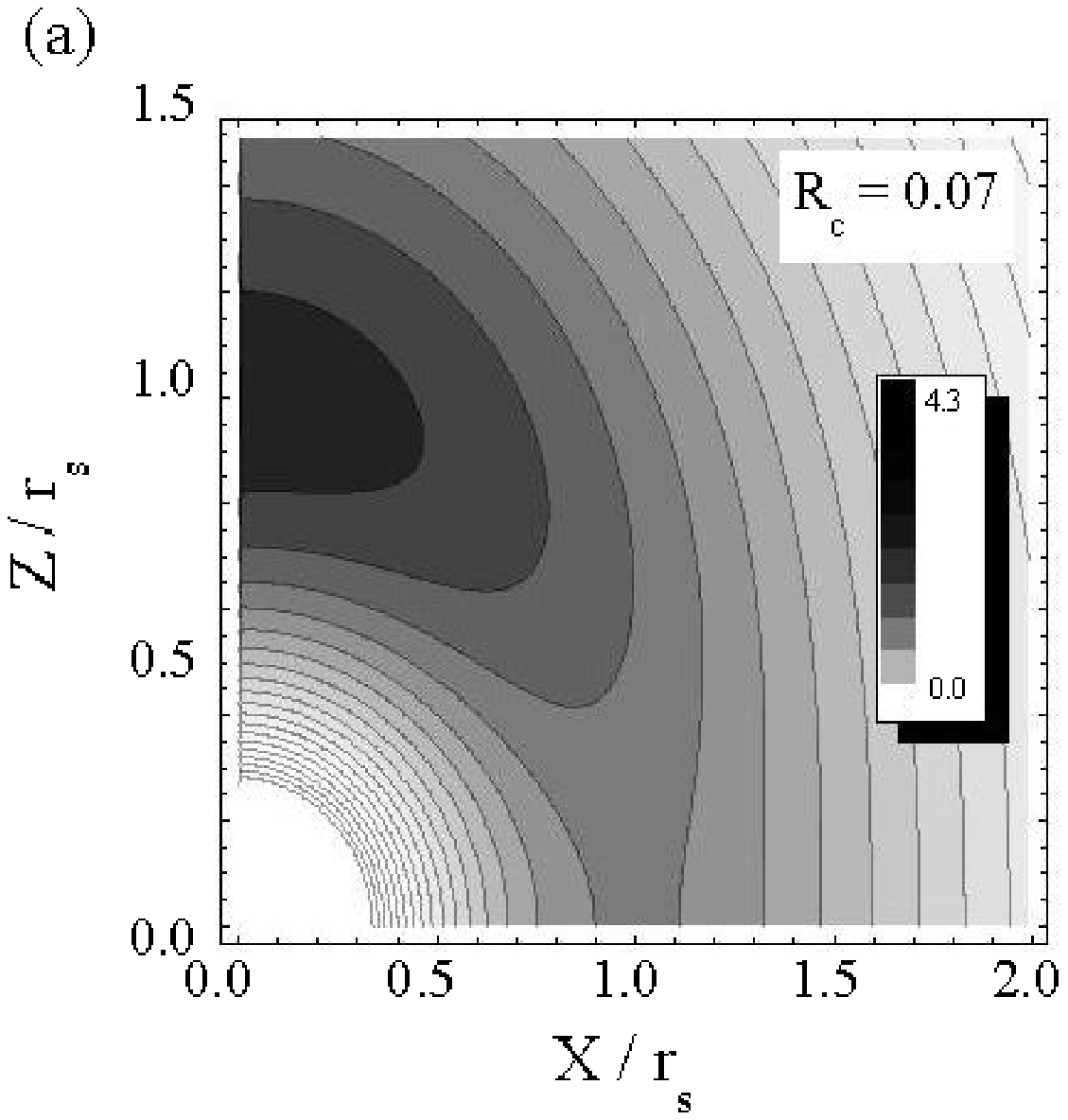}} &
      \resizebox{50mm}{!}{\includegraphics{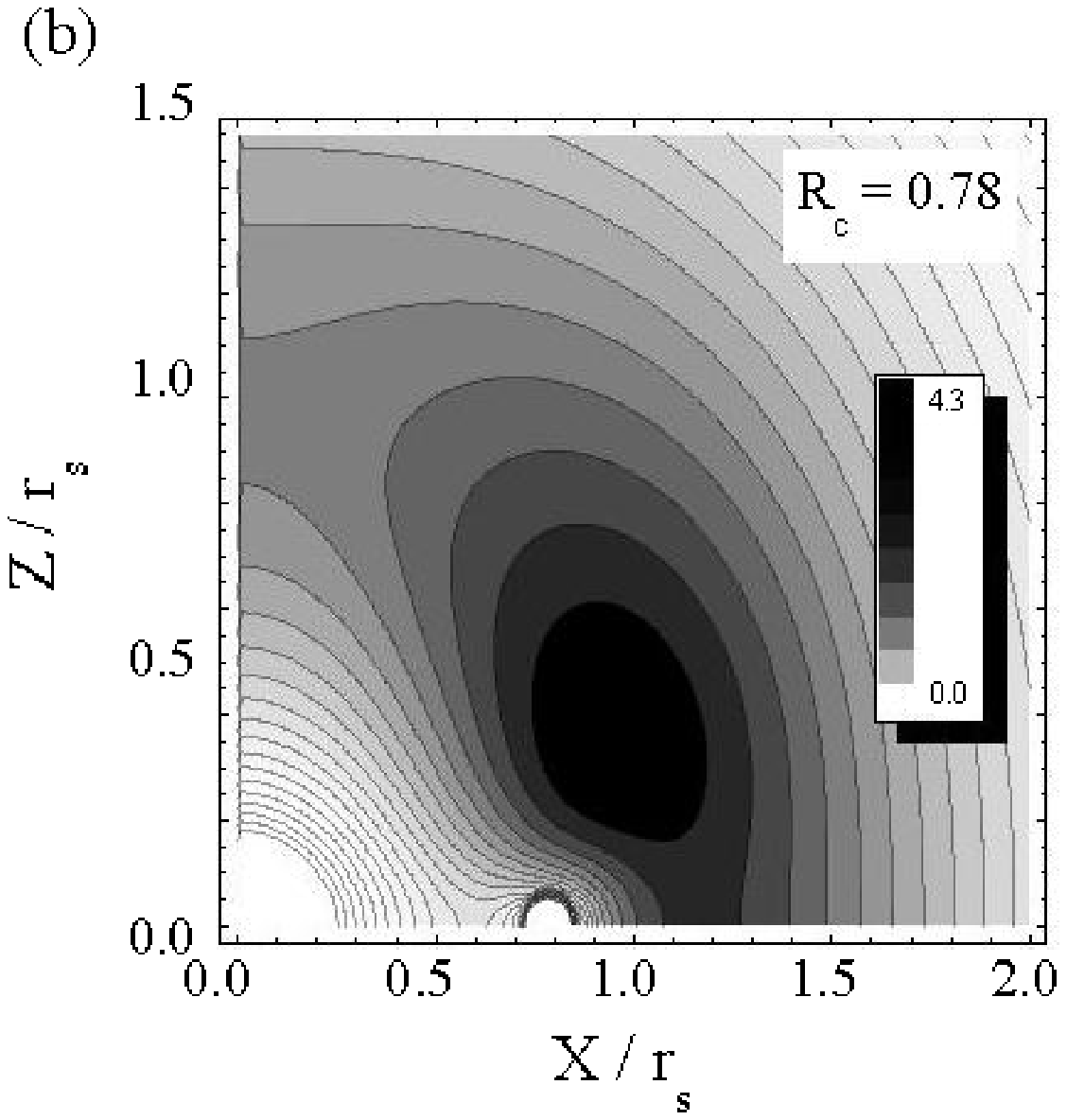}} &
      \resizebox{50mm}{!}{\includegraphics{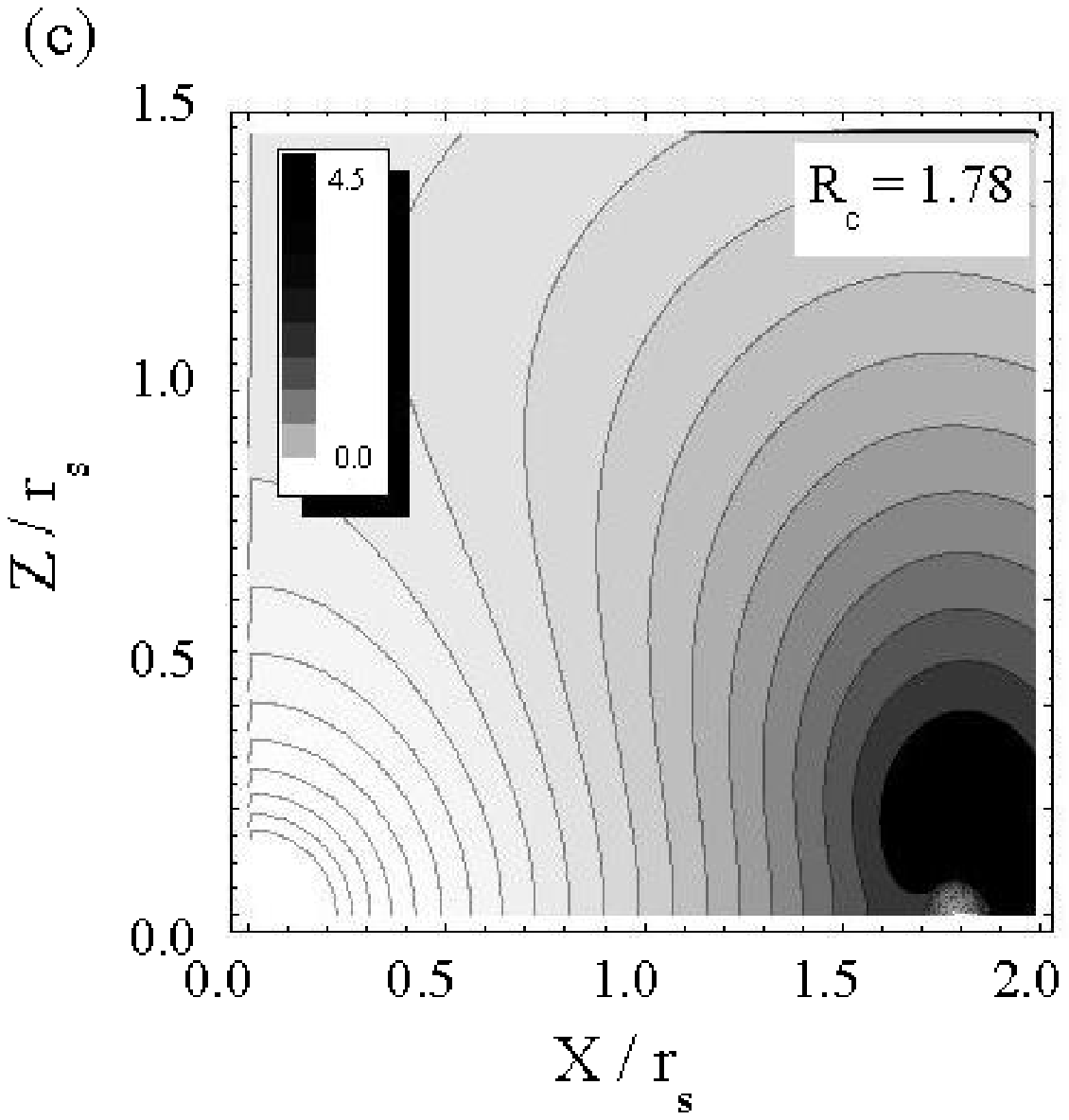}}\\
    \end{tabular}
    \caption{\label{C2}Contours of Kretchmann invariant, $\log_{10}{\cal I}^{(4)}$, corresponding to Fig.\ref{horizon_r0.1}.}
  \end{center}
\end{figure}

We next show the results of the homogeneous toroidal matter configurations.
Fig.\ref{horizon_r0.1} shows the two typical shapes of apparent horizons.
We also show the contours of ${\cal I}^{(4)}$ in Fig.\ref{C2}.
We set the ring radius of toroidal configurations as $R_r/r_s=0.1$ and search the sequence by changing the circle radius $R_c$.
When $R_c$ is less than $0.78r_s$, we find that only the $S^3$-apparent horizon (``common horizon" over the ring) exists. On the other hand, 
when $R_c$ is larger than $R_c=0.78r_s$, only the $S^1 \times S^2$ horizon (``ring horizon", hereafter) is observed. 
Unlike the cases of $\delta$-function matter distributions\cite{ida}, we could not find an example which shows both two horizons exist together.

We find that the value of ${\cal I}_{\rm max}$ appears at the outside of matter configuration as well as the spheroidal cases. 
Interestingly, ${\cal I}_{\rm max}$ is not hidden by the horizon when $R_c$ is larger $\left[\right.$see the case (c) of Fig.\ref{horizon_r0.1}$\left.\right]$.
This tendency is analogous to the spheroidal cases.
Therefore, if the ring matter shrinks itself to the ring, then a ``naked ring" (or naked di-ring) might be formed.

\begin{figure}[htbp] 
  \begin{center}
      \resizebox{7cm}{!}{\includegraphics{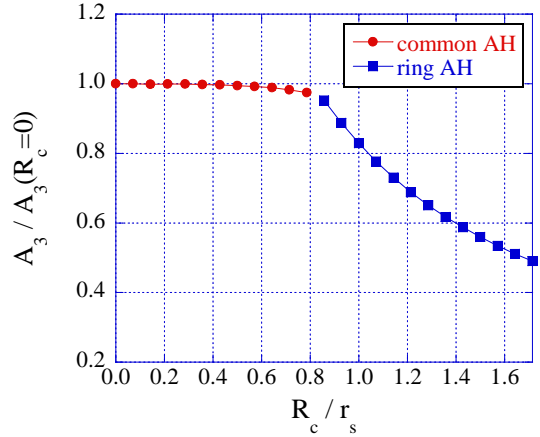}}
    \caption{\label{A2}The area of the apparent horizon $A_3$ for the toroidal matter distribution cases $R_c/r_c=0.1$. Plots are normalized by the area of spherical case ($R_c=0$). Two types of horizons do not exist simultaneously.
We see both horizons' area are smoothly connected at $R_c/r_s=0.78$, and both monotonically decrease with $R_c/r_s$.}
  \end{center}
\end{figure}

We show the surface area of the apparent horizons $A_3$ in Fig.\ref{A2}.
In Fig.\ref{A2}, typical two horizon monotonically decrease with $R_c/r_c$, the largest one is when the matter is in the spheroidal one ($R_c/r_c=0$).
We also observe that the common-horizon area is always larger than $S^1 \times S^2$ horizon area and two are smoothly connected in the plot.
If we took account the analogy of the thermodynamics of black-hole, this may suggest that if the black-ring evolves to shrink its circle radius then the ring horizon will switch to the common horizon at a certein radius.

\begin{figure}[htbp] 
  \begin{center}
      \resizebox{7cm}{!}{\includegraphics{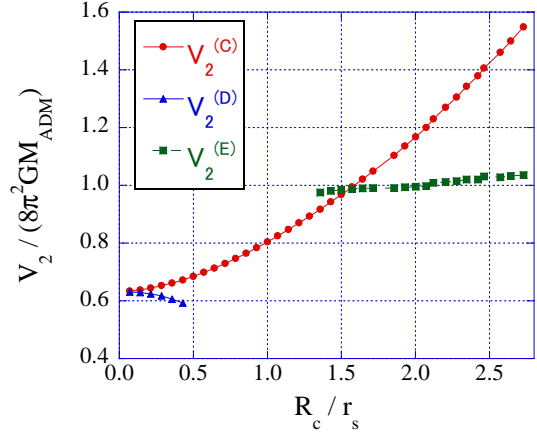}}
    \caption{\label{HH2}The ratio of the hyper-hoops $V_2$ to the mass $M_{ADM}$ are shown for the sequence of Fig.6. The ratio less than unity indicates that the validity of the hyper-hoop conjecture, Eq.(\ref{sufficient_condition}). We plot the hoops $V_2^{(C)}$, $V_2^{(D)}$ and $V_2^{(E)}$ where they exist. The horizon switches from the common horizon to ring horizon at $R_c/r_s=0.78$. Figure shows that the hoop $V_2^{(C)}$ and hoop $V_2^{(D)}$ represent the hyper-hoop conjecture for common apparent horizons properly, while $V_2^{(E)}$ is not for the ring horizon.}
  \end{center}
\end{figure}

Fig.\ref{HH2} shows the hyper-hoop $V_2^{(C)}$, $V_2^{(D)}$, and $V_2^{(E)}$ for these matter configurations.
We plot the points where we found hyper-hoops.
We note that $R_c/r_s=0.78$ is the switching radius from the common apparent horizon to the ring apparent horizon, and that $V_2^{(C)}$ and $V_2^{(D)}$ are sufficiently smaller than unity if there is a common apparent horizon.
Therefore, Eq.(\ref{sufficient_condition}) is satisfied for the formation of common horizon.
On the other hand, for the ring horizon, we should consider the hoop $V_2^{(E)}$ in Eq.(\ref{sufficient_condition}).
In Fig.\ref{HH2}, in the region $R_c/r_s > 0.78$, $V_2^{(E)}$ exists only a part in this region and becomes larger than unity.
Hence, for $S^1 \times S^2$ apparent horizon, the hyper-hoop conjecture, (\ref{sufficient_condition}), is not a proper indicator.
We conclude that the hyper-hoop conjecture, (\ref{sufficient_condition}), is only consistent with the formation of common horizon in toroidal case as far as our definition of the hyper-hoop is concerned.
\clearpage
\section{Summary and Future Works}
With the purpose of investigating the fully relativistic dynamics of
five-dimensional black-objects,
we constructed sequences of initial data and discussed the
formation of the apparent horizons,
the area of the horizons, and the validity of the hoop conjecture.

We modeled the matter in two cases; non-rotating homogeneous
spheroidal shape,
and toroidal shape under the momentarily static assumption.
Two models are still highly simplified ones, but the results
are well agreed with the previous semi-analytic works (both with 3+1  
and 4+1 dimensional
studies) and we also obtained new sequences for finite-sized matter  
rings.

We examined the so-called {\it hyper-hoop} conjecuture, where
the hoop is the {\it area} in 4+1 dimensional version.
We defined the hyper-hoop $V_2$ as it satisfies $\delta V_2=0$, and
searched the hoops numerically.

For the spheroidal matter cases, our results are simply the  
extensions of the previous studies.
The horizon is not formed when the matter is highly thin-shaped,
the hyper-hoop conjecture using our $V_2$ is properly satisfied,
and the maximum of the Kretchmann invariant ${\cal I}_{max}$ appears  
at the outside of the matter.
As was shown in the 3+1 dimensional case \cite{nakamura, shapiro}, this suggests also the
formation of a naked singularity when we start time evolution from this  
initial data.

While for the toroidal matter cases, both horizons and  
hoops can take two
topologies, $S^3$ and $S^1 \times S^2$, so that we considered both.
The apparent horizon is observed to switch from the common horizon  
($S^3$)
to the ring horizon ($S^1 \times S^2$) at a certain circle radius, and
the former satisfies the hyper-hoop conjecture, while the latter is not.
This is somewhat plausible, since the hoop conjecture was initially
proposed only for the 3+1 dimensional gravity where only the simply-connected black-hole is allowed.

From the area of the horizon and from the thermo-dynamical analogy of  
black holes,
we might predict the dynamical feature of the black-ring.  As we showed  
in Fig.\ref{A2},
the common horizon has larger area than the ring horizon, so that if  
the dynamics
proceed to shrink its circle radius, then a black-ring will naturally  
switch to a single black-hole.  However, if the local gravity is strong, then the ring  
might begin
collapsing to a ring singularity, that might produce also to the  
formation of
`naked ring'  since
${\cal I}_{max}$ appears on the outside of the ring (actually  
double-rings,
both on the top and the bottom of matter may be formed) for a certain initial configuration.
This is still a speculation and requires full dynamics in the future.

The initial-data sequences we showed here do not include rotations in  
matter and
space-time, which is one of our next subjects.  We now began studying the generalization of our models
including the
known exact solutions, that we hope to report elsewhere soon.
We are also developing our code to follow the dynamical processes in
five-dimensional spacetime, there we expect to show the validity of
the cosmic censorship  and hyper-hoop conjecture for various  
black objects.

\section*{Acknowledgments}
The numerical calculations were carried out on Altix3700 BX2 at YITP in Kyoto University.

\section*{}


\end{document}